\documentclass[prx,aps,twocolumn,reprint,superscriptaddress,floatfix,longbibliography]{revtex4-2}

\usepackage{appendix}

\usepackage{epsfig}
\usepackage{xfrac}
\usepackage{amsmath}
\usepackage{booktabs}
\usepackage{array}

\usepackage{microtype}
\usepackage{wasysym}
\usepackage[varg]{txfonts}
\usepackage[percent]{overpic}
\usepackage{mathrsfs}
\usepackage{braket}
\usepackage{titlesec}
\usepackage{bm}
\usepackage{comment}
\usepackage{verbatim}
\usepackage{cancel}
\usepackage{mathtools}

\usepackage{array}
\newcommand{\PreserveBackslash}[1]{\let\temp=\\#1\let\\=\temp}
\newcolumntype{C}[1]{>{\PreserveBackslash\centering}p{#1}}
\newcolumntype{R}[1]{>{\PreserveBackslash\raggedleft}p{#1}}
\newcolumntype{L}[1]{>{\PreserveBackslash\raggedright}p{#1}}

\usepackage{color}
\usepackage[colorlinks,bookmarks=false,citecolor=blue,linkcolor=blue,urlcolor=blue,pdfstartview=FitH]{hyperref}

\definecolor{darkred}{rgb}{0.7,0.0,0.0}

\definecolor{darkblue}{rgb}{0,0.02,0.45}

\definecolor{darkgreen}{rgb}{0.02,0.45,0.0}

\definecolor{violet}{rgb}{0.8,0.2,0.6}

\usepackage{times}
\usepackage{amsmath}
\usepackage{amssymb}
\usepackage{graphicx}
\usepackage{subfigure}
\usepackage{pict2e}
\usepackage{multirow}
\usepackage{float}

\newcommand{\be}{\begin{equation}}
\newcommand{\ee}{\end{equation}}
\newcommand{\bea}{\begin{eqnarray}}
\newcommand{\eea}{\end{eqnarray}}
\newcommand{\sbe}{\small\begin{equation}}
\newcommand{\see}{\end{equation}\normalsize}
\newcommand{\sbea}{\small\begin{eqnarray}}
\newcommand{\seea}{\end{eqnarray}\normalsize}
\graphicspath{{Figures/}}

\def\bs{\boldsymbol}

\def\mc{\mathcal}

\newcommand{\mat}{Cu$_2$OSO$_4$}

\graphicspath{{}{./Figures/}}

\begin{document}

\title{Frustration relief and reorientation transition in the kagome-like dolerophanite Cu$_2$OSO$_4$}

\author{Amelia Panther}
\affiliation{Department of Physics, Loughborough University, Loughborough LE11 3TU, United Kingdom}

\author{Alexander A. Tsirlin}
\affiliation{Experimental Physics VI, Center for Electronic Correlations and Magnetism, University of Augsburg, 86159 Augsburg, Germany}
\affiliation{Felix Bloch Institute for Solid-State Physics, Leipzig University, 04103 Leipzig, Germany}

\author{Ioannis Rousochatzakis}
\affiliation{Department of Physics, Loughborough University, Loughborough LE11 3TU, United Kingdom}

\begin{abstract}
We present a theoretical study of dolerophanite {\mat}, a layered kagome-like spin-$\frac12$ magnetic insulator that can be described either as a system of chains coupled through dimers or as a kagome lattice where every third spin is replaced by a ferromagnetic spin dimer. Building on insights from \textit{ab initio} calculations, classical numerical minimizations, and semiclassical expansions, we arrive at a minimal microscopic description that accounts for the experimental data reported so far, including the nature of the magnetic order, the reported spin length, and the observed anisotropy. The latter arises by a peculiar competition between the antisymmetric (Dzyaloshinskii–Moriya) and the symmetric part of the exchange anisotropy, which gives rise to a two-step re-orientation process involving two successive continuous phase transitions. Our work uncovers mechanisms stabilizing canted ferrimagnetic order in kagome systems, and highlights strong magnetic anisotropy in the presence of dissimilar magnetic orbitals on crystallographically nonequivalent Cu sites. We also show how these anisotropy terms affect the spin-wave spectrum and how they can be tracked experimentally. 
\end{abstract}

\maketitle

\section{Introduction}\label{sec:Introduction}

The search for quantum spin liquids remains at the forefront of condensed matter and quantum magnetism research for many decades~\cite{Diep,HFMBook,Ramirez94,Richter2004,PatrickLee2008,Balents2010,Norman2016,Savary2017,Zhou2017}, since P. W. Anderson's first proposal of the resonating valence bond idea~\cite{Anderson1973, FazekasAnderson74,Anderson1987,LiangDoucotAnderson88}.  
One of the main classes of candidate materials that have been explored intensively over the years are transition-metal compounds with dominant isotropic interactions and strong geometric frustration, most notably the 3D pyrochlores and quasi 2D kagome magnets~\cite{Diep,HFMBook,Gingras2014,Mendels2016}. 
In such materials, spins can evade magnetic ordering down to very low temperatures, despite their strong exchange interactions, due to the proliferation of infinite competing states at low energy scales. In conjunction with low dimensionality and low spin quantum number $S$, this competition magnifies the effects of quantum-mechanical fluctuations, and opens the possibility for unconventional classical and non-classical states, including spin nematics, valence bond crystals, and gapped and gapless spin liquids~\cite{Diep,HFMBook,Ramirez94,Richter2004,Balents2010,Norman2016,Savary2017,Zhou2017}.

While the above ingredients -- frustration, low dimensionality and low spin -- are present in many of the candidate materials that have been identified and characterized over the years, the broader consensus is that these same ingredients are also responsible for the high sensitivity to perturbations that are inevitably present in these compounds, such as longer-range interactions, magnetic anisotropy and various forms of randomness.  
Understanding the role of these perturbations is therefore crucial for explaining experimental data and for developing a useful phenomenology for the search of unconventional states. 

\begin{figure*}[!t]
\includegraphics[width=0.95\linewidth]{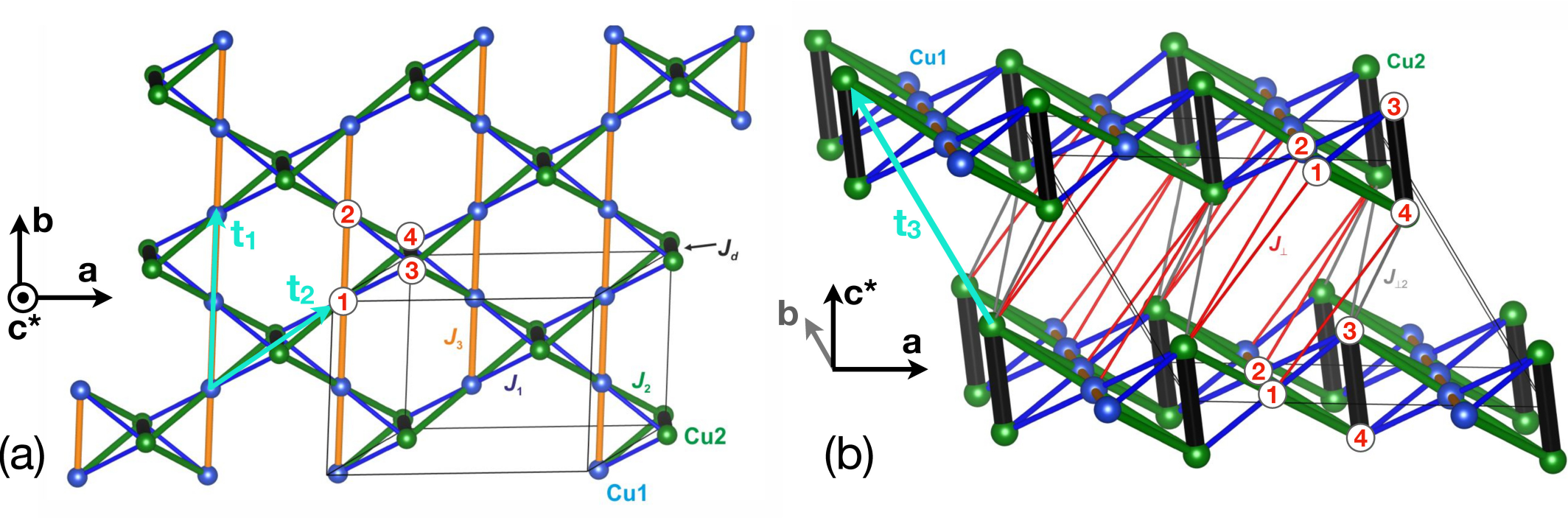}
\caption{(a) $(a,b)$ view of a single layer of {\mat}. Cu1 and Cu2 atoms are shown by blue and green spheres, respectively. The exchange coupling $J_3$ (orange) couples NN Cu1 atoms along  chains on the ${\bf b}$ axis, $J_d$ (thicker black lines) couples NN Cu2 atoms, and $J_1$ and $J_2$ couple neighbouring Cu1 and Cu2 atoms (blue and green lines, respectively).  The vectors ${\bf t}_1={\bf b}$, ${\bf t}_2=({\bf a}+{\bf b})/2$ and ${\bf t}_3={\bf c}$ are primitive translation vectors, and the numbers 1-4 (in red) label the four atoms of the basis. 
(b) Two layers of {\mat}, showing the topology of the couplings $J_\perp$ and $J_{\perp2}$ (red and grey lines, respectively).}\label{fig:Structure}
\end{figure*}

Here we present a theoretical study of dolerophanite {\mat}, a highly-frustrated, quasi-2D, kagome-like magnet, where magnetic anisotropy plays a key role. {\mat} has been known as a mineral since the early 1960s~\cite{OG1963}. Its structural characterization has been presented in Refs.~\cite{Effenberger1985,Ginting1994,Martens2003}, but its magnetic behaviour was explored only recently~\cite{Takahashi2012, Zhao2019, Favre2020}. The geometry of a given layer in {\mat} can be thought of as a system of chains joined through dimers or, alternatively, as a kagome lattice where every third spin is replaced by a dimer (Fig.~\ref{fig:Structure}). The material consists of two symmetry-nonequivalent Cu$^{2+}$ ions denoted by Cu1 and Cu2; the ions form the kagome lattice where every third vertex is occupied by a Cu2 dimer, and Cu1 ions form chains along the {\bf b}-axis. As it turns out, the two spin-$\frac12$ moments that make up a given dimer are so strongly ferromagnetically coupled that it would also be possible to consider an effective mixed-spin kagome model where one third of the sites is represented by $S=1$ moments. 

At first glance, experimental data for {\mat} suggest only a weak magnetic anisotropy revealed by the Curie-Weiss temperatures ($\Theta_{\rm CW}$) of $-71$\,K, $-75$\,K and $-70$\,K measured along the ${\bf a}$, ${\bf b}$, and ${\bf c^\ast}$ directions, respectively~\cite{Favre2020}. While the negative $\Theta_{\rm CW}$ values would suggest predominantly antiferromagnetic couplings, the magnetic ground state of {\mat} determined by neutron diffraction below $T_N\!=\!20$~K is best described as a canted ferrimagnet~\cite{Favre2020}. Spins adopt a uniform coplanar configuration with a 120$^\circ$-like structure in the $ab$-plane. The ferromagnetically coupled Cu2 spins point along the {\bf b}-axis, resulting in an uncompensated total magnetic moment of $\sim0.23(3)\mu_B$/Cu along this direction~\cite{Favre2020,Zhao2019}. This value is remarkably close to the classical value of $\mu_B/4$ expected for the coplanar 120$^\circ$ spin arrangement in a mixed-spin kagome magnet. Moreover, the ferromagnetic alignment of adjacent kagome layers implies that the total moments from different layers do not compensate each other.

Reflecting on these preliminary reports raises some important questions, which we seek to answer in our work. First and foremost, we wish to construct a minimal microscopic model that accounts for the experimental data~\cite{Takahashi2012,Zhao2019,Favre2020}, and elucidate to what extent frustration of the kagome layers can account for the physics of {\mat}. We also wish to examine the origin of the uncompensated  magnetic moment, understand why it is pinned to a specific direction in the crystal structure, and what determines the preferred plane for the coplanar spins.
 
Ultimately, in addressing these questions we also aim to understand what sets {\mat} apart from the conventional kagome physics. At first glance, such a deviation is not  unexpected, due to the structural peculiarities of {\mat}, in particular, the presence of four Cu sites in the unit cell, instead of three, and the absence of a three-fold symmetry and the inequivalence between the  nearest-neighbour (NN) exchange paths. However, a closer analysis reveals that, despite these differences, the Heisenberg model of {\mat} remains highly frustrated, with an infinite ground state manifold that is similar, e.g., to that of the kagome francisites~\cite{Rousochatzakis2015}, and undistorted kagome magnets. Yet, the observation of a robust low-temperature ordered state with an almost classical value of the spin lengths~\cite{Zhao2019,Favre2020} indicates minimal quantum fluctuations despite the frustration. 

This brings us to the next main focus of this work, which is to uncover the key role played by the magnetic anisotropy in this compound. As we show below, the lifting of the infinite ground state degeneracy of the isotropic model, and the selection of the observed, almost classical order, arises from a non-trivial interplay of the antisymmetric part of the exchange anisotropy, i.e., the Dzyaloshinskii–Moriya (DM) interactions~\cite{Dzyaloshinsky1958, Moriya1960}, and the symmetric (traceless) part, which we shall denote by ${\bf T}$ in this work. According to our Density Functional Theory (DFT) calculations, the relevant DM vectors feature significant components along the crystallographic ${\bf c}^\ast$ axis, and much weaker components in the $ab$ plane. This hierarchy explains the selection of the observed uniform coplanar state and the pinning of the spins on the $ab$ plane, but fails to reproduce the direction of the uncompensated moment in that plane, as the in-plane DM components select the ${\bf a}$ and not the (observed) ${\bf b}$ axis. The inclusion of the symmetric part of the exchange anisotropy resolves this final puzzle and leads to a reorientation of the total moment along ${\bf b}$, provided this anisotropy has large enough strength to override the effect of the in-plane DM components, and that it has the appropriate sign. Our DFT calculations for one of the most relevant bonds confirm explicitly that such an anisotropy is present in {\mat}. We further show that the re-orientation of the total moment proceeds via a two-step process involving an intermediate phase, and that the spin gap depends very sensitively on the interplay between the in-plane DM components and the symmetric  part of the exchange anisotropy.

Having identified the right minimal model, our final aim is to carry out self-consistent checks but also make predictions for further experiments, which, in turn, will allow for more quantitative estimates of the microscopic coupling parameters in this complex, multi-sublattice system. To that end, we shall present a study of the response of {\mat} under a magnetic field along the three crystallographic directions, and uncover the presence of appreciable transverse magnetization which can be probed by magnetic torque measurements. We shall also study the effect of quantum fluctuations and confirm that the reduction of the spin length is indeed very small, in agreement with experiment. Finally, we shall present the magnon band structure from linear spin-wave calculations. 

The remainder of the paper is organized as follows. In Sec.~\ref{sec:Structure} we highlight the main aspects of the crystal structure of {\mat}, including the global symmetries, the different coordination of the two types of Cu ions, and their lattice topology. In Sec.~\ref{sec:AbInitioI} we discuss details of DFT calculations (Sec.~\ref{sec:AbInitioMethods}) and the resulting electronic structure (Sec.~\ref{sec:DOS}), which sheds light into the nature of the magnetic orbitals of the two types of Cu sites in {\mat}. In Sec.~\ref{sec:Jmodel} we present our study of the effect of isotropic Heisenberg interactions, the predicted hierarchy of energy scales from DFT (Sec.~\ref{sec:AbInitioResultsJ}), the resulting picture for the infinite classical ground state manifold and its close similarity to the 2D KHAFM (Sec.~\ref{sec:IsotropicGSs}), and the effect of the interlayer couplings (Sec.~\ref{sec:Interlayer}). In Sec.~\ref{sec:IncludingDM} we present what happens when we include the microscopic DM interactions. This includes an analysis of the constraints imposed by symmetry on the DM vectors (Sec.~\ref{sec:DM-sym-constraints}), their numerical values from DFT (Sec.~\ref{sec:DFT-DMvectors}), the resulting picture for the structure of the selected ground states (Sec.~\ref{sec:JplusDM-GSs}), and a comparison to experiments. In Sec.~\ref{sec:IncludingTbb} we incorporate the symmetric part of the exchange anisotropy tensors, discuss their constraints from symmetry (Sec.~\ref{sec:T-Sym-constraints}), identify particular elements of these tensors that can turn the total moment along the observed direction (Sec.~\ref{sec:Tmatrix-relevant}), and present DFT results on one of the relevant bonds that corroborate this picture (Sec.~\ref{sec:DFT-Tbb}). In Sec.~\ref{sec:reorientationtransition} we show, more generally, that the rotation of the total moment proceeds via a two-step reorientation transition, and discuss the evolution of the spin gap through this transition. In Sec.~\ref{sec:OtherAspects} we study some additional topics and make predictions for further experiments: The response of {\mat} under a magnetic field along the three crystallographic directions (Sec.~\ref{sec:MvsB}), the effect of quantum fluctuations at the quadratic level (Sec,~\ref{sec:LSWT}), the associated reduction of the spin length (Sec.~\ref{sec:SpinLength}) and the magnon band structure (Sec.~\ref{sec:magnons}). In Sec.~\ref{sec:Conclusions} we give our conclusions and a broader perspective of our work. Finally, we include two appendices (App.~\ref{app:StructuralAspects} and \ref{appendix:CWT}) which contain technical details and auxiliary information.

\begin{figure*}[!t]
\includegraphics{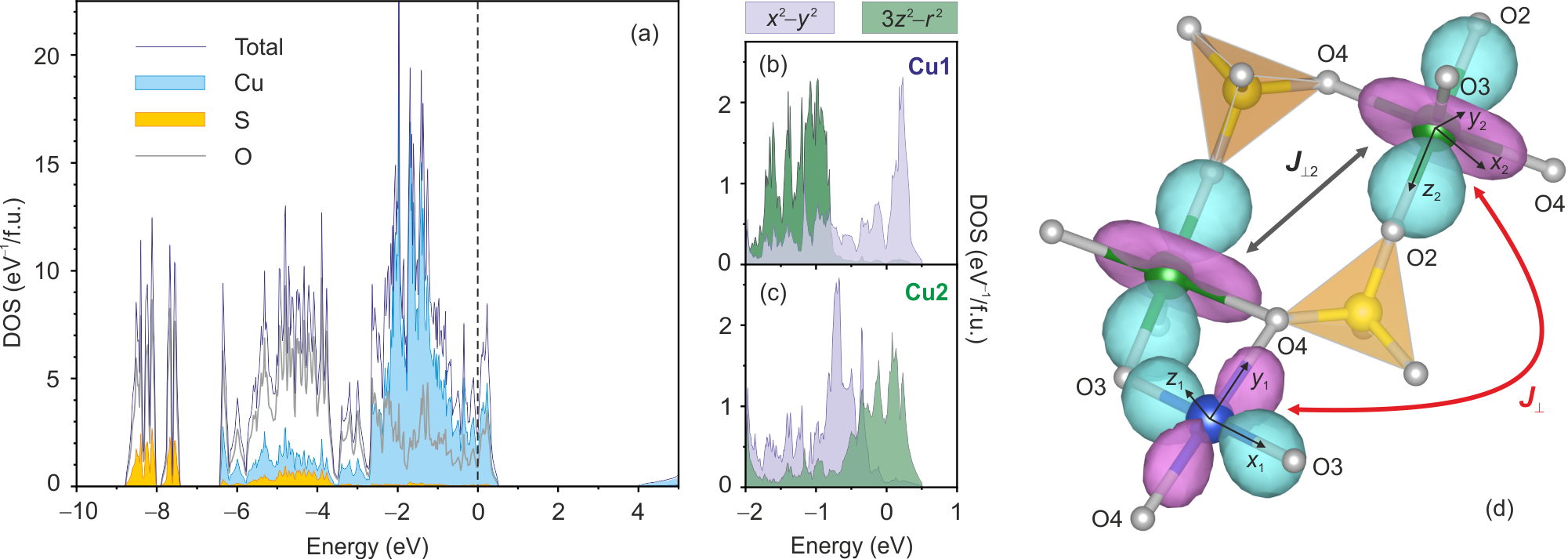}
\caption{(a) Electronic density of states (DOS) for {\mat} calculated on the PBE level shows predominant Cu $3d$ states near the Fermi level, which is denoted by the dashed line at zero energy. (b,c) Orbital-resolved DOS indicates different symmetry of the magnetic orbitals for the Cu1 and Cu2 atoms. (d) Magnetic orbitals visualized using Wannier functions and superexchange pathways for the interlayer couplings $J_{\perp}$ and $J_{\perp2}$. Local coordinate frames for the Cu1 and Cu2 atoms are also shown. The notation of oxygen positions follows Ref.~\cite{Effenberger1985}.}\label{fig:superexchange}
\end{figure*}

\section{Main structural and symmetry aspects}\label{sec:Structure}
Figure~\ref{fig:Structure}\,(a) shows the kagome-like structure of each $ab$ layer of {\mat}, and Fig.~\ref{fig:Structure}\,(b) shows the way such layers arrange along the ${\bf c}^\ast$ axis.  There are two symmetry nonequivalent Cu$^{2+}$ sites in {\mat}, denoted by Cu1 and Cu2. The nearest-neighbour (NN) Cu2 sites form dimers,  whereas Cu1 sites form 1D chains along the ${\bf b}$ axis. Thus, each layer can be thought of as either a system of chains joined through dimers or as a kagome lattice with every third spin replaced by a dimer. The overall spin lattice can be described in terms of the underlying monoclinic Bravais lattice, plus a basis of four atoms, two Cu1 and two Cu2, see Fig.~\ref{fig:Structure}.

{\mat} crystallizes in a monoclinic crystal system belonging to the space group $C2/m$~\cite{OG1963}, where the monoclinic ${\bf c}$-axis forms an angle of $\beta=122.34^\circ$ with the ${\bf a}$-axis. The vector ${\bf c}^\ast$ is perpendicular to the $ab$ plane, as shown below
\be\label{eqfig:newaxis}
\includegraphics[width=1.75in]{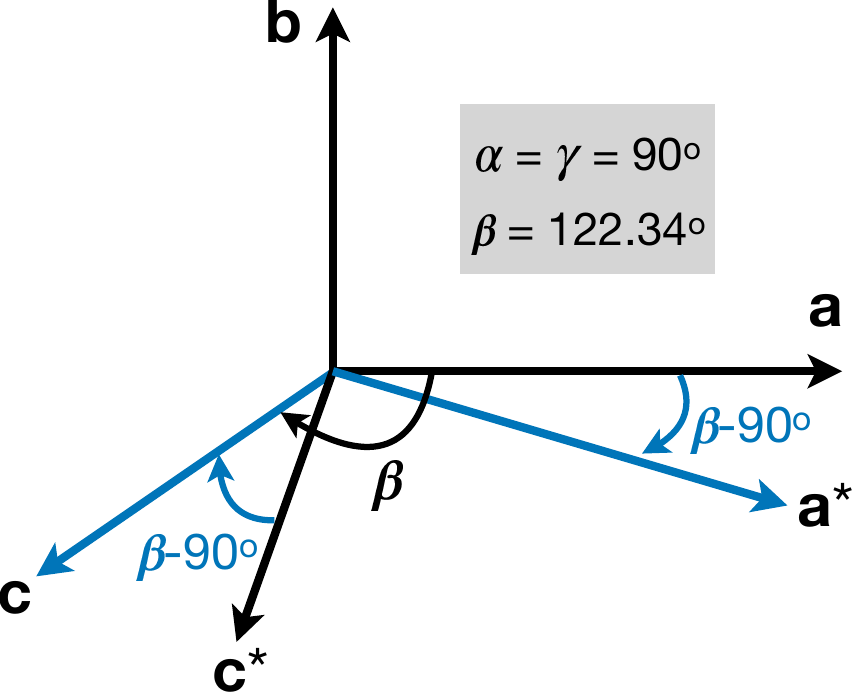}
\ee 
Apart from primitive translations, the $C2/m$ symmetry group includes:  i) inversion center in the middle of the Cu2 dimers;  ii) inversion center in the middle of the $J_{\perp,2}$ bond; iii) two-fold rotation axis $C_{2{\bf b}}$ passing through the middle of the Cu2 dimers; iv) screw axis along the Cu1 chain corresponding to a non-primitive translation by ${\bf b}/2$ followed by a two-fold rotation $C_{2{\bf b}}$ around the ${\bf b}$ axis; and v) $ac^\ast$ mirror plane that goes through the midpoints of NN Cu1 (or $J_3$) bonds and includes the Cu2 dimers. 
The experimental magnetic structure does not break any of these symmetries (it belongs to the irrep $\Gamma_1$ of the $C2/m$ space group at zero momentum~\cite{Favre2020}).

Another important aspect of the {\mat} structure concerns the local coordination of the two types of Cu$^{2+}$ ions, which affects the choice of the magnetic orbital. The Cu1 atoms are surrounded by four oxygen atoms with the Cu--O distances of $2\times 1.882$\,$\AA$ (Cu1--O3) and $2\times 2.070$\,$\AA$ (Cu1--O4), see Fig.~\ref{fig:superexchange}(d). Two further oxygen atoms are located 2.526\,$\AA$ away from copper and belong to the SO$_4$ groups. Therefore, Cu1 has the 4+2 oxygen coordination with the square plaquette of four nearest oxygen atoms obtained by an axial elongation of the CuO$_6$ octahedron. By contrast, Cu2 features two short Cu2--O2 distances of 1.907\,$\AA$ followed by three further oxygens at 2.000\,$\AA$ (Cu2--O3) and $2\times 2.155$\,$\AA$ (Cu2--O4), resulting in the 2+3 trigonal-bipyramidal coordination. This structural configuration gives rise to a rather unusual coupling regime that we discuss in the following.

\section{Density-functional theory \& electronic structure}\label{sec:AbInitioI}

\subsection{Methods}\label{sec:AbInitioMethods}
Density-functional (DFT) band-structure calculations were performed in the \texttt{FPLO}~\cite{fplo} and \texttt{VASP} codes~\cite{Kresse1996, Kresse1996_2} using the Perdew-Burke-Ernzerhof (PBE) flavor of the exchange-correlation potential \cite{Perdew1996}. Correlation effects in the Cu $3d$ shell were taken into account on the mean-field DFT+$U$ level with the on-site Coulomb repulsion $U_d=8.5$\,eV (\texttt{FPLO}) or 9.5\,eV (\texttt{VASP}), Hund’s coupling $J_d$=1 eV, and atomic limit for the double-counting correction \cite{Mazurenko2014, Janson2014}. Note that $U_d$ is applied to the Cu $3d$ orbitals, hence there is usually a difference in the optimal $U_d$ value depending on the DFT code and its basis set \cite{Tsirlin2010}. Experimental structural parameters from Ref. \cite{Effenberger1985} were used in all calculations.  
The various microscopic spin interactions, including the Heisenberg exchange parameters presented below in Sec.~\ref{sec:Jmodel}), the DM vectors (Sec.~\ref{sec:IncludingDM}) and the symmetric exchange anisotropy ${\bf T}_3$ (Sec.~\ref{sec:IncludingTbb}), have been obtained by a mapping procedure~\cite{Xiang2011,Tsirlin2014} from total energies of magnetically ordered states. These energies were converged on a $k$ mesh with 64 points in the first Brillouin zone for the supercells with 64 atoms (double the crystallographic unit cell of \mat).

Each interaction parameter is derived from total energies of four spin configurations as follows,
\be
J_{12}=\frac{E_{\uparrow\uparrow}-E_{\uparrow\downarrow}-E_{\downarrow\uparrow}{+E_{\downarrow\downarrow}}}{4S^2}
\ee
where $S=\frac12$ and $\uparrow\uparrow$, $\uparrow\downarrow$, $\downarrow\uparrow$, $\downarrow\downarrow$ denote spin arrangements on the interacting atoms 1 and 2 while spins on all other atoms are kept fixed to a globally ferromagnetic configuration. The $\alpha\beta$ component of the interaction tensor $\mathbb J$ is obtained in a similar way by setting the spin on the atom 1 along $\alpha$ and the spin on the atom 2 along $\beta$, whereas the spins on all other atoms are kept along $\gamma$ perpendicular to both $\alpha$ and $\beta$. When $\alpha\neq\beta$, the resulting component is a combination of the DM interaction ($D^{\gamma}$) and off-diagonal symmetric anisotropy ($T^{\alpha\beta}$) that are separated as
\be
\epsilon_{\alpha\beta\gamma}D^{\gamma}=(\mathbb J^{\alpha\beta}-\mathbb J^{\beta\alpha})/2
\ee\be
T^{\alpha\beta}=(\mathbb J^{\alpha\beta}+\mathbb J^{\beta\alpha})/2,
\ee

where $\epsilon_{\alpha\beta\gamma}$ is the Levi-Civita symbol. The \texttt{FPLO} values are for the Heisenberg exchange parameters only, because noncollinear spin configurations are not implemented in this DFT code. Additional data related to the DFT results of this work can be found in Ref.~\cite{dft-data}.

\subsection{Electronic structure}\label{sec:DOS}
The uncorrelated (PBE) band structure of {\mat} shows predominantly oxygen states below $-3$\,eV followed by Cu $3d$ bands that extend to slightly above the Fermi level, see Fig.~\ref{fig:superexchange}(a). The band structure is metallic because correlations in the Cu $3d$ shell have not been included in the calculation. The states near the Fermi level show contributions from both the $d_{x^2-y^2}$ and $d_{3z^2-r^2}$ orbitals, although only one of them should be eventually half-filled and magnetic in the $3d^9$ Cu$^{2+}$ ion. Choosing the $x_1$ and $y_1$ local coordinate axes along the shorter Cu1--O3 and Cu1--O4 bonds within the square plaquette results in a separation of the two orbitals, with the $d_{x^2-y^2}$ states lying at higher energies than the $d_{3z^2-r^2}$ ones, see Fig.~\ref{fig:superexchange}(b). Therefore, $d_{x^2-y^2}$ is the likely magnetic orbital for Cu1. A similar separation is possible in the case of Cu2, but here the local $z_2$ axis has to be directed along the two shortest Cu2--O2 bonds. The $d_{3z^2-r^2}$ states are then higher in energy than the $d_{x^2-y^2}$ ones, and $d_{3z^2-r^2}$ is the likely magnetic orbital for Cu2, see Fig.~\ref{fig:superexchange}(c).

This analysis is verified by DFT+$U$ calculations that produce an insulating solution. Correlations stabilize the orbital state with the unpaired electron on the $d_{x^2-y^2}$ orbital for Cu1 and on the $d_{3z^2-r^2}$ orbital for Cu2. This is similar to the high-temperature phase of another kagome-like mineral, volborthite~\cite{Janson2010,Yoshida2012}, but different from the majority of other Cu-based kagome magnets, including francisite~\cite{Rousochatzakis2015} where both Cu sites feature $d_{x^2-y^2}$ as the magnetic orbital.
\vspace{0.9cm}

\begin{table*}[!t]
\caption{Microscopic magnetic parameters of Cu$_2$OSO$_4$: leading exchange couplings $J_i$ (in K) obtained from \texttt{FPLO} ($U_d$\,=8.5 eV) and \texttt{VASP} ($U_d$\,=9.5 eV), and corresponding DM vectors ${\bf D}_{ij}$ (\texttt{VASP}), calculated for each bond type. Here, the labelling of the two sites (`site $i$' and `site $j$') of each bond follows the notation of Table~\ref{tab:StructuralAspects} in App.~\ref{app:StructuralAspects}. Note that inversion symmetry forbids the DM interactions on the $J_d$ and $J_{\perp2}$ bonds.}
\begin{ruledtabular}
\begin{tabular}{llllllll}
bond type & site~$i$ & site~$j$& ${\bf r}_{j}-{\bf r}_i$~($abc^\ast$ frame, in ${\AA}$)  &$|{\bf r}_j-{\bf r}_i|$ (${\AA}$)& $J_{ij}^{\,\texttt{FPLO}}$ (K) & $J_{ij}^{\,\texttt{VASP}}$ (K) & ${\bf D}_{ij}$~  ($abc^\ast$~frame, in K)    \\
\hline
$J_d$ (Cu2--Cu2)  		&Cu22&Cu22p				&  $(0.4322, 0, -2.8166)$		&2.849&$-199$	& $-168$  	& ${\bf D}_d=0$ \\
$J_1$ (Cu1--Cu2)  		&Cu11&Cu22				&  $(2.1264, 1.5798, 1.4083)$	&3.000& 20		&48 		& ${\bf D}_1=(43,-39, -38)$ \\
$J_2$ (Cu1--Cu2) 		&Cu11&Cu22p				&  $(2.5586, 1.5798, -1.4083)$	&3.320&66		&56 		& ${\bf D}_2=(-31, 25, -30)$ \\
$J_3$ (Cu1--Cu1) 		&Cu11&Cu14				& $(0,3.1595,0)$  			&3.159&74		& 72 		& ${\bf D}_3=(15, 0, 81)$ \\
$J_{\perp2}$ (Cu2--Cu2) 	&Cu22&Cu23			 	& $(1.0307, -3.1595, 3.6375)$	&4.927&15		&15 		&   ${\bf D}_{\perp,2}=0$  \\
$J_{\perp}$ (Cu1--Cu2)	&Cu11&Cu23+{\bf b}/2    		& $(3.1572, 1.5798, 5.0458)$ 	&6.158&28 		&26 		& ${\bf D}_\perp=(5, -1, -3)$
\end{tabular}
\end{ruledtabular}
\label{tab:abinitio}
\end{table*}

\begin{figure}[!h]
\includegraphics{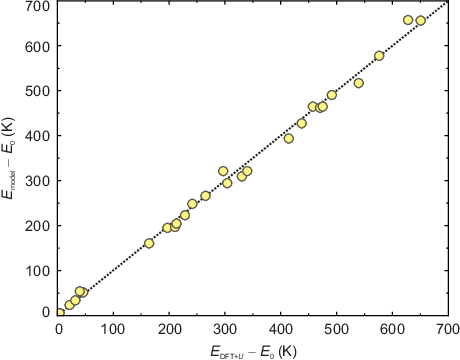}
\caption{\label{fig:energies}
Comparison of total energies for different spin configurations as evaluated by DFT+$U$ and obtained from the minimal model with six exchange couplings listed in Table~\ref{tab:abinitio}. Both sets of energies are given relative to the reference energy $E_0$ that corresponds to the nonmagnetic state.}
\end{figure}

\section{Isotropic model}\label{sec:Jmodel}
We begin our analysis with an investigation of the isotropic Heisenberg spin Hamiltonian 
\be
\mc{H}_{\text{iso}}=\sum\nolimits_{i<j}J_{ij}{\bf S}_i\cdot{\bf S}_j\,,
\ee
where $J_{ij}$ denotes the exchange parameter between spin-1/2 operators ${\bf S}_i$ and ${\bf S}_j$ at Cu sites $i$ and $j$, respectively.
\vspace{1cm}

\subsection{DFT results for Heisenberg exchange parameters}\label{sec:AbInitioResultsJ}
Exchange couplings for all Cu--Cu pairs with the distances of less than 7\,$\AA$ were calculated using supercells doubled either in the $ab$ plane or along the ${\bf c}$ direction. Two complementary DFT codes produced consistent results except for the coupling $J_1$ that is twice larger in \texttt{VASP} compared to \texttt{FPLO}. Both sets of parameters will be considered in the following. The leading Heisenberg exchange couplings are listed in Table \ref{tab:abinitio} and are also indicated in Fig.~\ref{fig:Structure}. Further exchange couplings can be neglected in the minimal model because they are below 10\,K. Fig.~\ref{fig:energies} shows that the six leading couplings considered in this work allow a good description of total energies for different spin configurations in \mat.

The coupling on the Cu2 dimers, which is denoted by $J_d$, sets the dominant energy scale and is ferromagnetic. The NN coupling on the Cu1 chains, which is denoted by $J_3$, is antiferromagnetic and gives the second strongest energy scale. The couplings $J_1$ and $J_2$ between Cu1 and Cu2 ions are also antiferromagnetic and compete with $J_3$. Finally, there are two types of interlayer interactions, $J_{\perp}$ and $J_{\perp2}$, which couple two Cu1 atoms to a single Cu2 atom, and two Cu2 atoms to a single Cu2 atom, respectively. The predominance of these couplings over other possible  superexchange pathways between the kagome layers can be explained by the presence of bridging SO$_4$ tetrahedra, as shown in panel (d) of Fig.~\ref{fig:superexchange}. The interlayer couplings are weaker than $J_2$ and $J_3$, but overall significant. Here, $J_{\perp2}$ is non-frustrated and at first glance would couple the layers antiferromagnetically. On the other hand, $J_{\perp}$ forms triangular loops with $J_3$ in the same way as $J_1$ does. The coupling $J_{\perp}$ would not be satisfied in the scenario of simple antiferromagnetic order imposed by $J_{\perp2}$. In the experimental magnetic structure, the $J_{\perp2}-J_{\perp2}-J_3$ triangles end up having the same type of canted order as the $J_1-J_1-J_3$ triangles, so $J_{\perp2}$ may act as an additional driving force of the canted order, see detailed analysis in Sec.~\ref{sec:IsotropicGSs}.

Let us now compare the experimentally measured values of the Curie-Weiss temperature $\Theta_{\text{CW}}$ to the values obtained from the DFT Heisenberg couplings of Table~\ref{tab:abinitio} and the expression (see discussion in App.~\ref{appendix:CWT})
\be\label{eq:ThetaCW}
\Theta_{\text{CW}} = -\frac{1}{8} [J_d+2(J_3+J_{\perp,2})+4(J_1+J_2+J_\perp)]\,,
\ee
where the sign convention of $\Theta_{\text{CW}}$ corresponds to the Curie-Weiss approximation for the susceptibility $\chi=\frac{C}{T-\Theta_{\text{CW}}}$, and $C$ is the Curie constant. 
The resulting values are provided in the first four rows of the second column of Table~\ref{tab:models}, where, for comparison, we give the values corresponding to \texttt{VASP} and \texttt{FPLO}, as well as the values without (2D) and with (3D) the interlayer couplings ($J_\perp$ and $J_{\perp,2}$) included. The experimentally reported values are $-68$\,K for a powder sample~\cite{Zhao2019}, and $-71$\,K, $-75$\,K and $-70$\,K along the ${\bf a}$, ${\bf b}$, and ${\bf c^\ast}$ axes,  respectively, for a single crystal~\cite{Favre2020}). 
The comparison shows that: i) including the interlayer couplings gives a more satisfactory agreement, and ii) the \texttt{VASP} values give a better agreement compared to \texttt{FPLO}.

\subsection{The isotropic model of a single layer: Highly-frustrated magnetism \& similarity to the KHAFM}\label{sec:IsotropicGSs}

The isotropic limit of a single layer of {\mat} features an infinite number of classical ground states, which are closely related to the ones of the ideal KHAFM~\cite{Elser1989,Harris1992,HuseRutenberg1992,Chalker1992}. To see this, we follow a cluster minimization method (see, e.g., Ref.~\cite{LyonsKaplan1964}) and examine the four-site unit cell of the structure.

\subsubsection{The building block of classical ground states} 
Let us first rewrite the total Heisenberg energy of a single layer, $\mc{H}_{\text{iso,1}}$, as a sum over contributions from unit cells,
\be\label{eq:UCell2}
\includegraphics[width=1.25in]{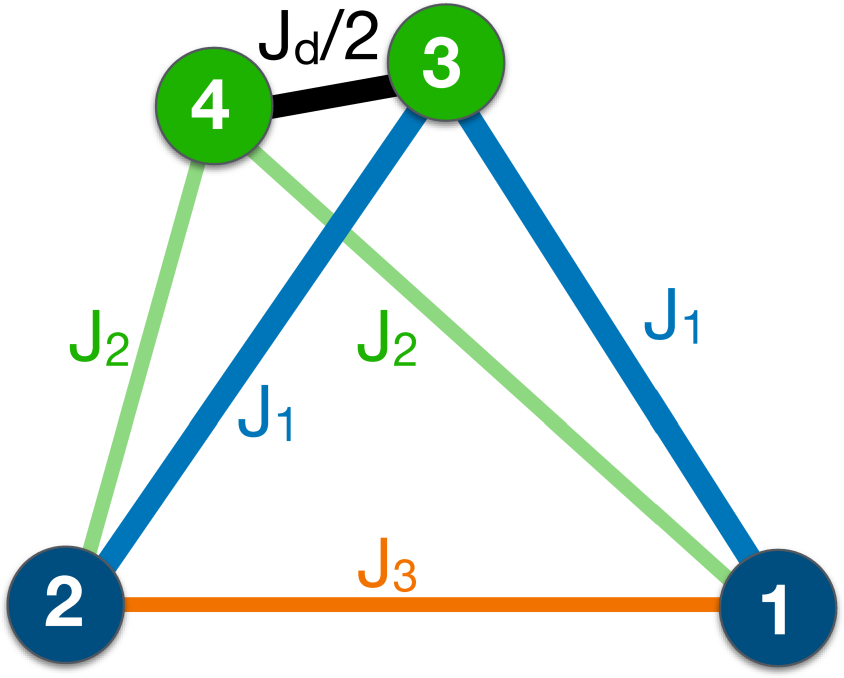}
\ee 
where $J_d$ is replaced by $J_d/2$ because this bond is shared by two cells. We have, 
\be\label{eq:Hisoblocks1}
\mc{H}_{\text{iso}}= \sum\nolimits_{\parbox{0.25in}{\epsfig{file=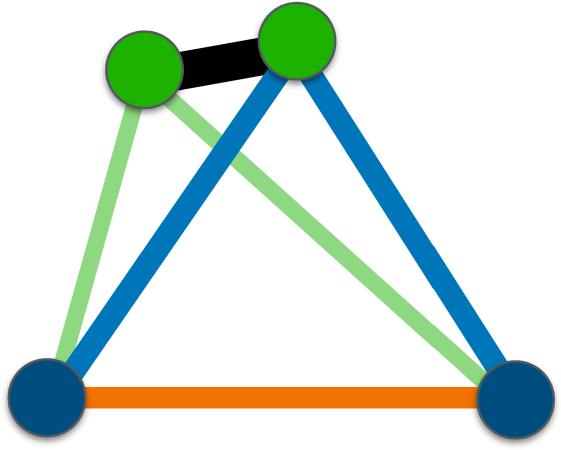,width=0.2in,clip=}}} \!\!\!\mc{H}_{\text{iso},\parbox{0.25in}{\epsfig{file=UCell,width=0.2in,clip=}}},
\ee
where
\be\label{eq:Hisoblocks2}
\mc{H}_{\text{iso},\parbox{0.25in}{\epsfig{file=UCell,width=0.25in,clip=}}}
\!\!\!= J_3 {\bf S}_1\cdot{\bf S}_2+(J_1{\bf S}_3+J_2{\bf S}_4)\cdot({\bf S}_1+{\bf S}_2)+\frac{J_d}{2}{\bf S}_3\cdot{\bf S}_4\,.
\ee
As is turns out, in the minimum energy configuration, the spins ${\bf S}_3$ and ${\bf S}_4$ point along the same direction and the energy reduces to that of an `isosceles Heisenberg triangle' with competing couplings $J_3$ and $J_1+J_2$, 
\be
\mc{H}'_{\text{iso},\parbox{0.25in}{\epsfig{file=UCell,width=0.25in,clip=}}}
= J_3 {\bf S}_1\cdot{\bf S}_2+(J_1+J_2) {\bf S}_3\cdot({\bf S}_1+{\bf S}_2)+J_d S^2/2\,,
\ee
where $S$ is the classical spin length. 
The minimum energy state takes the schematic form
\be\label{eq:UCellState}
\includegraphics[width=1.95in]{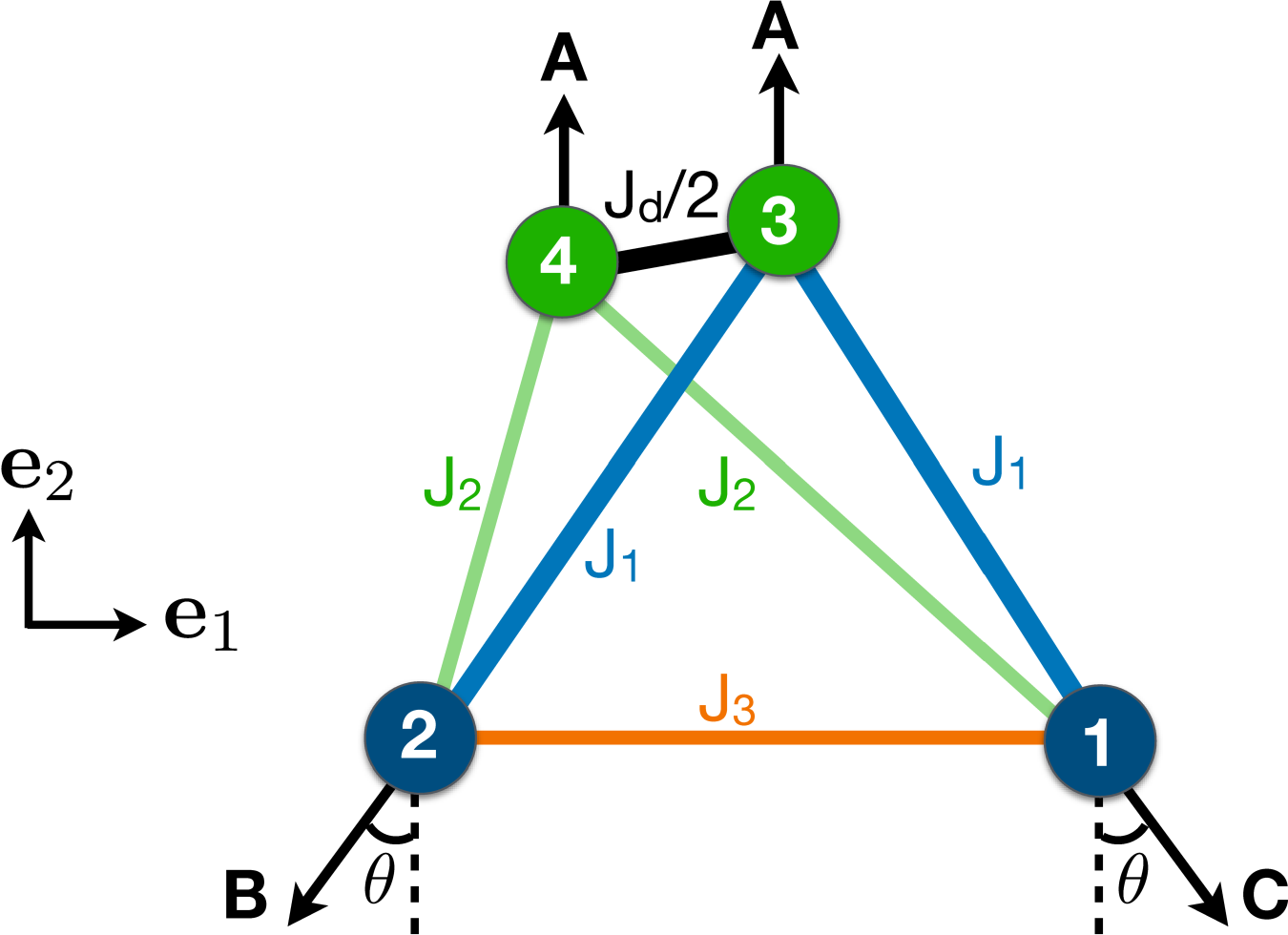}
\ee 
or, more explicitly, 
\be\label{eq:theta1}
\renewcommand\arraystretch{1.2}
\begin{array}{l}
{\bf S}_3/S={\bf S}_4/S={\bf e}_2 \equiv {\bf A},
\\
{\bf S}_1/S=\sin\theta~{\bf e}_1 -\cos\theta~{\bf e}_2 \equiv {\bf B},
\\
{\bf S}_2/S=-\sin\theta~{\bf e}_1-\cos\theta~{\bf e}_2 \equiv {\bf C},
\end{array}
\ee
where ${\bf e}_1$ and ${\bf e}_2$ are any pair of orthogonal axes, and the angle $\theta$ can be found by minimizing the energy $E=J_3\cos(2\theta)-2(J_1+J_2)\cos\theta+J_d S^2/2$, which gives 
\be\label{eq:theta1b}
\theta=
\begin{cases}
    \cos^{-1}\bigg(\frac{J_1+J_2}{2J_3}\bigg),& \text{if } \frac{|J_3|}{|J_1+J_2|} \geq \frac{1}{2} \\
    0,              & \text{otherwise}
\end{cases}
\ee
The values of $\theta$ predicted from \texttt{FPLO} and \texttt{VASP} are 54.5$^\circ$ and 43.8$^\circ$, respectively. From these, the former is closer to the value of $60^\circ$ realized in the `equilateral triangle' limit of $J_1+J_2=J_3$, which seems to be the relevant region for {\mat}~\cite{Favre2020}. 

The total magnetic moment per Cu can be obtained by
\be\label{eq:ClassMperCu}
{\bf m}/(g\mu_B)=\frac{S}{2}(1-\cos\theta)~{\bf e}_2\,. 
\ee
Taking $S=1/2$ and assuming $g=2$, we find $m\simeq 0.21\mu_B$ (\texttt{FPLO}) and $0.14\mu_B$ (\texttt{VASP}) per Cu. The former is only slightly lower than the value $0.23(3)\mu_B$ found experimentally~\cite{Zhao2019}. So, at the level of the isotropic model of a single layer, the \texttt{FPLO} values give a better agreement for the the length of the total moment. This comparison is premature however, since, as we show below (see also fourth column of Table~\ref{tab:models}), these numbers show significant variation when including the interlayer couplings and/or the anisotropies.

\begin{figure*}[!t]
\includegraphics[width=0.33\linewidth]{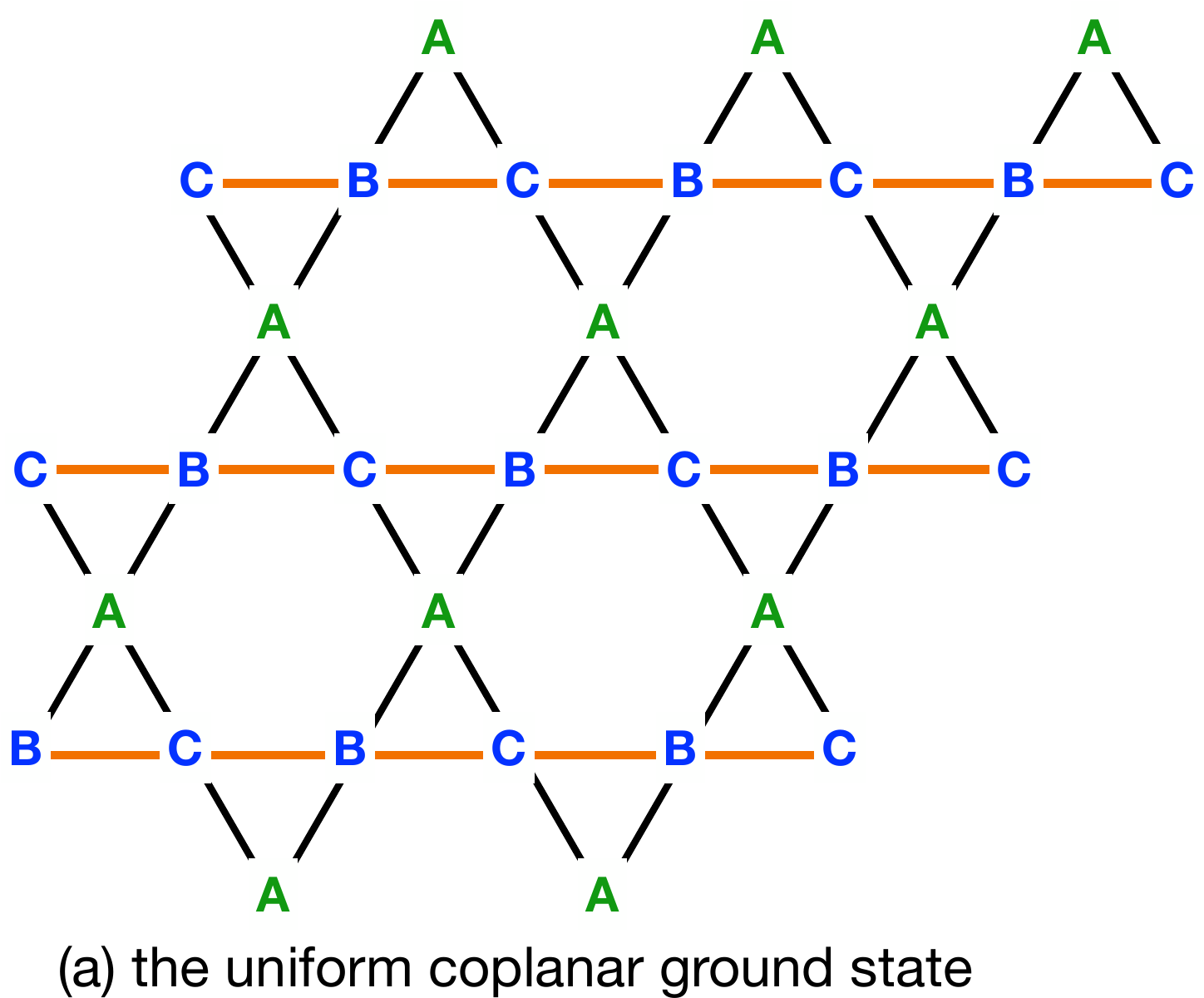}
\includegraphics[width=0.33\linewidth]{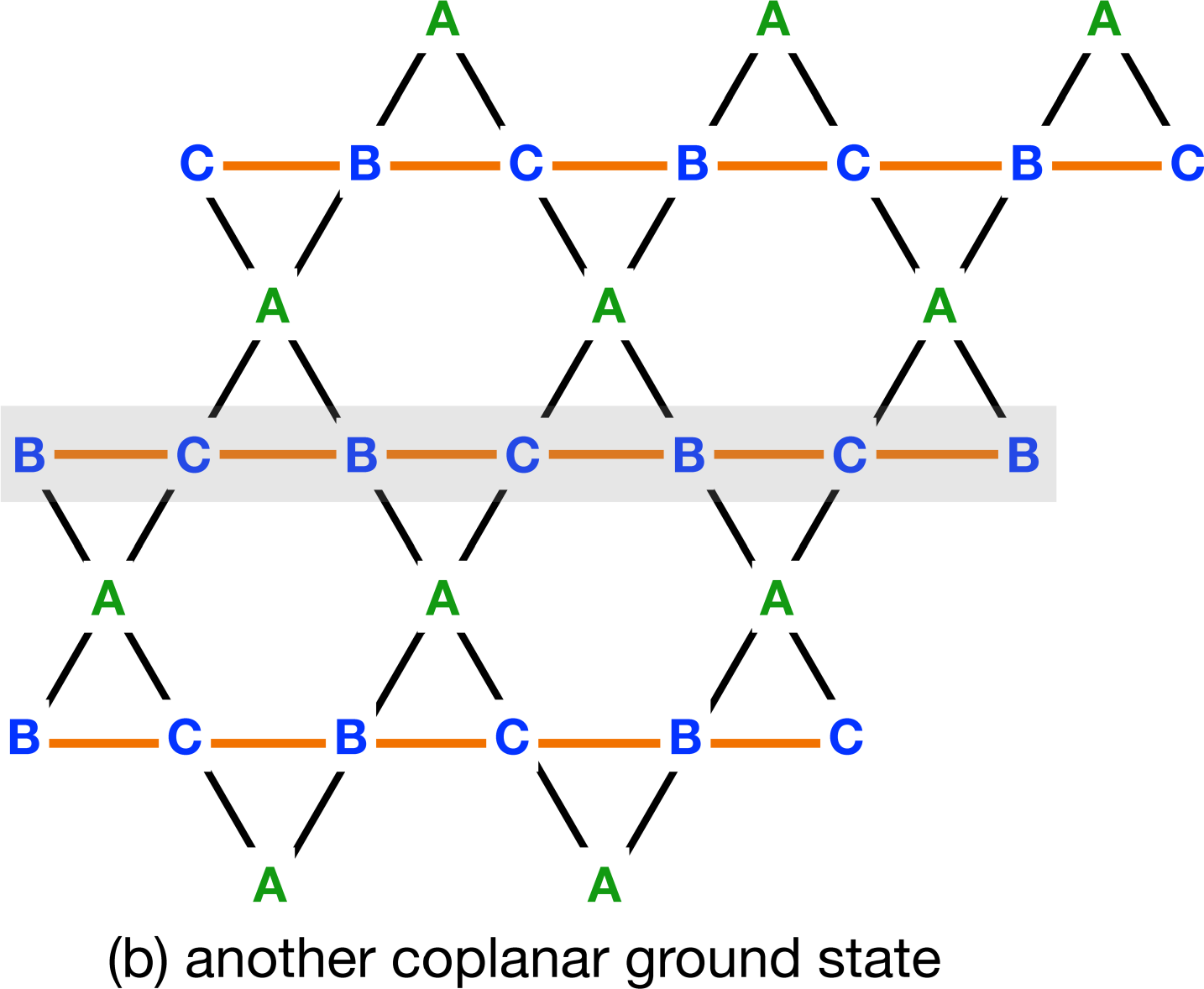}
\includegraphics[width=0.33\linewidth]{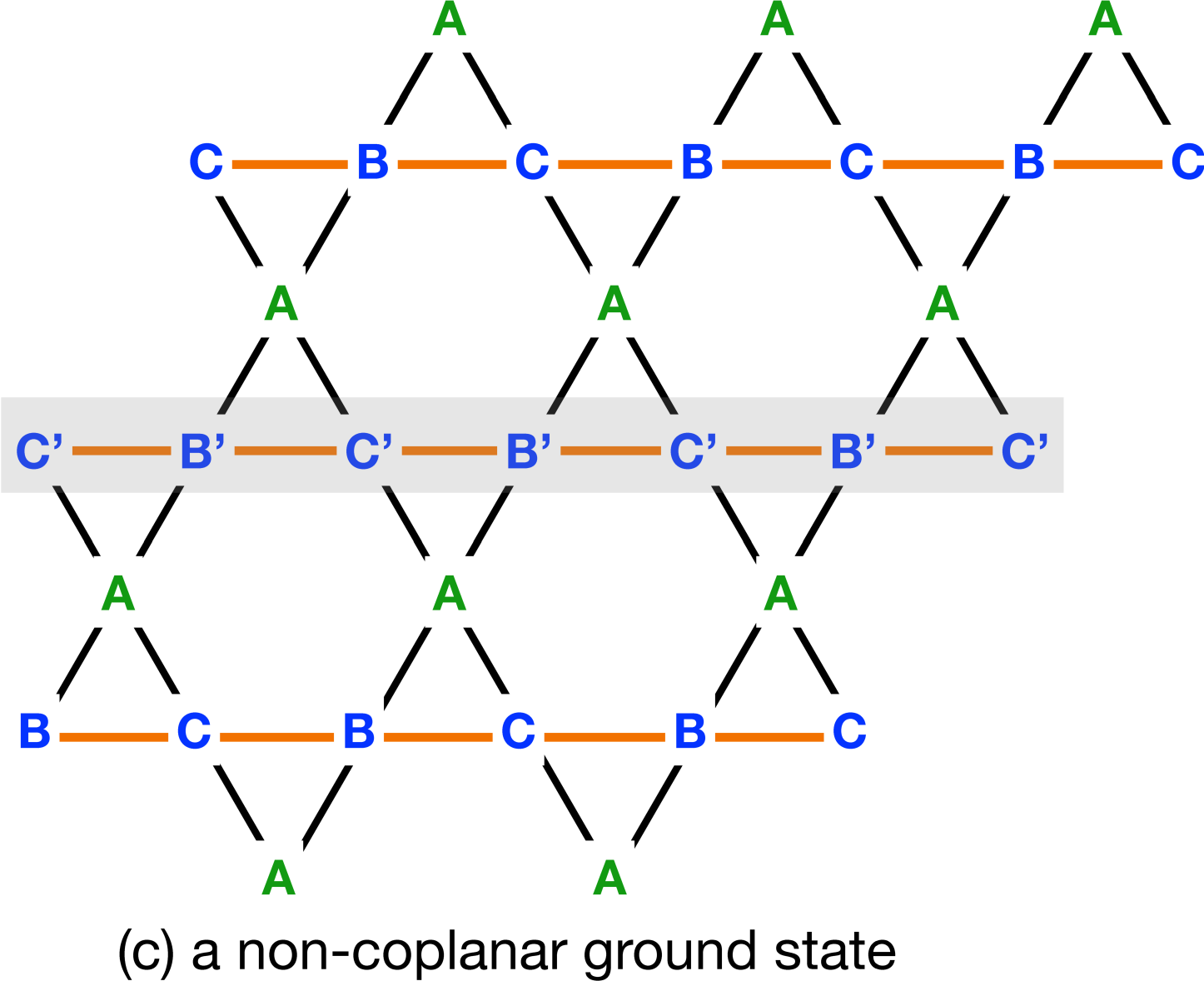}
\caption{Representative members of the classical ground state manifold of the isotropic model of a single layer of {\mat}, where Cu2 dimers are represented by one of the three types of vertices of the kagome ({\bf A} sites). 
({\bf a}) The uniform coplanar ABC state, where A, B and C denote the spin directions of Eq.~(\ref{eq:theta1}) of the building block of the lattice. 
({\bf b}) One of the $2^L$ coplanar ground states of a single layer (where $L$ is the number of horizontal rows of the layer). This state results from the one in ({\bf a}) by swapping {\bf B}$\leftrightarrow${\bf C} in one of the rows of the layer (shaded).  
({\bf c}) One of the non-coplanar ground states of a single layer. Here, this state results from the one in ({\bf a}), by rotating the spins belonging to a given row around the direction of {\bf A} by some arbitrary angle $\phi$, which effectively rotates {\bf B} and {\bf C} to some new directions {\bf B'} and {\bf C'}, respectively, out of the plane of {\bf A}, {\bf B} and {\bf C}.}\label{fig:ClassGSs}
\end{figure*}

\subsubsection{Tilling the `ABC' state onto the lattice} 
Having found the ground states of one building block of the Hamiltonian, we can now proceed to tile these on the full lattice. The resulting ground state manifold consists of infinite ground states, including an infinite subset of coplanar states and an infinite subset of non-coplanar states. Some representative members of this manifold are shown in Fig.~\ref{fig:ClassGSs}.

{\it Coplanar States --} The uniform coplanar state of Fig.~\ref{fig:ClassGSs}\,(a) is the simplest way to tile the ABC state of Eq.~(\ref{eq:theta1}) to the full lattice, and is also the one that has been observed experimentally~\cite{Favre2020}. This state has an SO(3) order parameter space, as there are two angles needed to fix the plane of {\bf A}, {\bf B} and {\bf C}, and an additional angle to fix the direction of (say) the {\bf A} sublattice within that plane.

However, this is not the only coplanar state. Indeed, starting from the uniform state of Fig.~\ref{fig:ClassGSs}\,(a), one can generate the coplanar state shown, e.g., in Fig.~\ref{fig:ClassGSs}\,(b) by choosing one of the $L$ horizontal rows of the lattice and swapping {\bf B}$\leftrightarrow${\bf C} along that row. Such an operation does not affect the energetics on the particular row of triangles, or on any other triangles of the lattice, and is therefore a ground state. Counting all possible sequences of swaps on horizontal chains gives $2^L$ such coplanar states. 

Similar swaps can be performed on non-horizontal chains as well. However, one {\it cannot} perform a sequence of swaps along chains of different direction, so the total degeneracy of coplanar states is $3\times 2^L$ (and not $2^{3L}$). 
The uniform state is a special member of this sub-extensive family of coplanar states.

\begin{table*}[!t]
\caption{Numerical values of various parameters based on DFT values of microscopic couplings. Different rows correspond to different levels of model descriptions. E.g., $\mc{H}_{\text{iso}}$ (\texttt{FPLO}, 2D) corresponds to a single layer of {\mat} with isotropic Heisenberg couplings only, and values given by \texttt{FPLO} code (see Table~\ref{tab:abinitio}), and $\mc{H}_{\text{iso}}+\mc{H}_{\text{DM}}$ (\texttt{VASP}, 3D) corresponds to the actual multilayer system, with isotropic and DM couplings (including $J_\perp$, $J_{\perp,2}$, ${\bf D}_\perp$ and ${\bf D}_\perp'$), and numerical values given by the \texttt{VASP} code. The second column gives the Curie-Weiss temperature $\Theta_{\text{CW}}$, which is given by Eq.~(\ref{eq:ThetaCW}) for the 2nd, 4th and 6th rows, and by the same equation but with $J_\perp$ and $J_{\perp,2}$ set to zero for the 1st, 3rd and 5th rows. For the last two rows, the three values given correspond to $\Theta_{\text{CW}}^{aa}$, $\Theta_{\text{CW}}^{bb}$ and $\Theta_{\text{CW}}^{c^{\ast}c^{\ast}}$, which are given by Eq.~(\ref{eq:ThetaCWaniso2}) in the appendix. The vector ${\bf m}$ denotes the total moment per Cu site and $m$ its length. For the meaning of the angles $\theta$, $\alpha$, $\psi$ and $\xi$ see Eq.~(\ref{eq:UniformState1})-(\ref{eq:UniformState3}).}
\begin{ruledtabular}\begin{tabular}{lcccclcc}
model & $\Theta_{\text{CW}}$~(K)   & direction of ${\bf m}$ ($abc^\ast$ frame) & $m/\mu_B$ ($g=2$)  & $\theta$ & $\alpha$  & $\psi$   & $\xi$ \\
\hline
$\mc{H}_{\text{iso}}$ ~(\texttt{FPLO}, 2D) & -36.6 &any &0.21 &54.5$^\circ$ &any&any&0\\ 
$\mc{H}_{\text{iso}}$ ~(\texttt{FPLO}, 3D) & -54.4  &any &0.11 &39.6$^\circ$&any&any&0\\ 
$\mc{H}_{\text{iso}}$ ~(\texttt{VASP}, 2D) & -49.0  &any &0.14 &43.8$^\circ$&any&any&0\\ 
$\mc{H}_{\text{iso}}$ ~(\texttt{VASP}, 3D) & -65.8  &any &0.05 &25.5$^\circ$&any&any&0\\ 
\toprule
$\mc{H}_{\text{iso}}+\mc{H}_{\text{DM}}$ ~(\texttt{VASP}, 2D)& -49.0  &(0.93, 0,  0.36) &0.22 &53.8$^\circ$&5.0$^\circ$ &90$^\circ$ & 11.1$^\circ$\\ 
$\mc{H}_{\text{iso}}+\mc{H}_{\text{DM}}$ ~(\texttt{VASP}, 3D)& -65.8 &(0.91, 0, 0.42) 	&0.22 &50.60$^\circ$ &2.60$^\circ$ & 90$^\circ$ &10.41$^\circ$\\ 
\toprule
$\mc{H}_{\text{iso}}+\mc{H}_{\text{DM}}+\mc{H}_{T_3}$~(\texttt{VASP}, 2D) & (-51.2, -47.8, -48.0) & (0, 1, 0) &0.21 &54.39$^\circ$ &3.62$^\circ$ &0&0\\
$\mc{H}_{\text{iso}}+\mc{H}_{\text{DM}}+\mc{H}_{T_3}$~(\texttt{VASP}, 3D) & (-68.0, -64.5, -64.8) & (0, 1, 0) &0.19 &51.13$^\circ$ &5.2$^\circ$ &0 & 0
\end{tabular}\end{ruledtabular}\label{tab:models}
\end{table*}

{\it Non-coplanar states --} The above operation of swapping {\bf B}$\leftrightarrow${\bf C} along a given row corresponds to a $\pi$-rotation of the spins along that row around the direction of the {\bf A} sublattice. 
This procedure can be generalized to rotations by {\it any} angle $\phi$. The spins along the given row will rotate around {\bf A} from {\bf B} and {\bf C} to some new directions {\bf B'} and {\bf C'}, which are out-of-the original ABC plane, see Fig.~\ref{fig:ClassGSs}\,(c). Given that all neighbours of that row of sites point along {\bf A}, a rotation around {\bf A} preserves the local structure and the relative angles necessitated by Eqs.~(\ref{eq:theta1}-\ref{eq:theta1b}) and does not alter the energy. In total, we can perform such a rotation by an angle $\phi_r$ on the $r$-th row of the lattice, and these angles are in general different, leading to an SO(2) degeneracy for each row (i.e., SO(2)$^L$ in total). Similar rotations can be performed along the non-horizontal chains as well. 

Besides the above non-coplanar ground states, one can envisage other ground states which could arise by performing a sequence of rotations along open or closed paths that are different from chains, provided we do not affect the energy of any triangle. While our numerical minimizations show evidence for such more complex ground states, it is not clear whether they belong to the above manifold of non-coplanar states that are generated by performing a sequence of rotations along chains. 
 
Summarizing the physics of the isotropic model on a single layer of {\mat}, we have found a classical ground state manifold with infinite coplanar and infinite non-coplanar ground states, which can all be built from the local 3-sublattice building block of Eq.~(\ref{eq:theta1}). The uniform coplanar state found experimentally is a special member of this manifold. 
While this state could potentially be the one that is selected by quantum-mechanical fluctuations, we anticipate (based on what happens, e.g., in the case of the KHAFM~\cite{Harris1992,Chalker1992,Chubukov1992,CCMkagome2011,Chernyshev2014kagomeXXZ,Ralko2018}) that the respective order-by-disorder energy scale is much smaller compared to the interlayer couplings or the anisotropy. And given that the latter may lift the classical ground state degeneracy already at the mean-field level, we shall disregard the order by disorder physics of the pure isotropic model, and proceed to investigate the effect of the interlayer couplings $J_\perp$ and $J_{\perp,2}$ and the anisotropy.

\subsection{Effect of interlayer couplings}\label{sec:Interlayer}

\subsubsection{Effect of $J_{\perp,2}$ alone}
Referring back to the right panel of Fig.~\ref{fig:Structure}, one sees that $J_{\perp,2}$ connects one Cu2 dimer from one layer to a Cu2 dimer on the next layer. So the main effect of this interaction is to fix the relative orientation of Cu2 dimers on successive layers, leading to an alternating {\bf A}, -{\bf A}, {\bf A}, $\cdots$ structure of the Cu2 dimers along the ${\bf c}^\ast$ axis.

However, $J_{\perp,2}$ {\it does not} lift the infinite degeneracy discussed above, and the system still features infinite coplanar and infinite non-coplanar ground states. 
For example, we can start from one of the coplanar states in a single layer and then `copy' its time-reversed version on the next layer, and continue in an alternating fashion with the next layers, so that we satisfy $J_{\perp,2}$. However, one can still swap {\bf B}$\leftrightarrow${\bf C} (or -{\bf B}$\leftrightarrow$-{\bf C}) along any Cu1 chain of any layer, without affecting the energy, leading again to a sub-extensive number of coplanar ground states. 
The situation for non-coplanar states can be explained in a similar way. 

Importantly, none of the members of the classical ground state manifold matches the one found experimentally, because even when the in-plane configuration is uniform, the relative orientation is AFM across the layers due to $J_{\perp,2}$. Hence, this interaction does not seem to be relevant for the explanation of the uncompensated moment in {\mat}.

\subsubsection{Effect of $J_{\perp}$ alone}
Let us now discuss the effect of $J_\perp$. According to the right panel of Fig.~\ref{fig:Structure}, $J_\perp$ connects a NN pair of Cu1 sites on one layer to a Cu2 site on the next layer. Effectively then, this coupling plays a role similar to that of $J_1$ and $J_2$, competing with $J_3$. At the mean-field level, this coupling gives rise to a parallel arrangement of the Cu2 dimers across different layers, in contrast to the antiparallel arrangement favoured by $J_{\perp,2}$. Additionally, $J_\perp$ preserves the ABC structure of Eq.~(\ref{eq:theta1}) of the building block of the lattice, but renormalizes the angle $\theta$ of Eq.~(\ref{eq:theta1b}) to 
\be\label{eq:theta1c}
\theta=
\begin{cases}
    \cos^{-1}\bigg(\frac{J_1+J_2+J_\perp}{2J_3}\bigg),& \text{if } \frac{J_3}{J_1+J_2+J_\perp} \geq \frac{1}{2} \\
    0,              & \text{otherwise}
\end{cases}
\ee
which arises from Eq.~(\ref{eq:theta1b}) by replacing 
\be
J_1+J_2\mapsto J_1+J_2+J_\perp\,,
\ee
reflecting the fact that $J_\perp$ plays a similar role with $J_1$ and $J_2$, as mentioned above.
The {\it ab initio} parameters of Table~\ref{tab:abinitio} give $\theta\simeq39.6^\circ$ (\texttt{FPLO}) and $25.5^\circ$ (\texttt{VASP}), which are further away from the 60$^\circ$ of the equilateral triangle case, compared to the ones without $J_\perp$, see comparison in fourth column of Table~\ref{tab:models}. 
In turn, these numbers give a total moment of $m\simeq0.11\mu_B$ (\texttt{FPLO}) and $0.05\mu_B$ (\texttt{VASP}) based on Eq.~(\ref{eq:ClassMperCu}), which are significantly reduced compared to the experimental values. 

Finally, using the same arguments as in the case of $J_{\perp,2}$ above, one finds that $J_\perp$ too fails to select a unique ground state, and the classical ground state manifold contains again infinite coplanar and infinite non-coplanar states. The only differences to the case of $J_{\perp,2}$ then are: i) the renormalization of the angle $\theta$, and ii) the fact that now the Cu2 dimers align ferromagnetically across the layers.

Importantly, the experimentally reported configuration is one of the members of the ground state manifold, although there is an appreciable deviation in the value of $\theta$. This suggests that either DFT overestimates the value of $J_\perp$, and/or that the anisotropic couplings can remedy this problem (see below).

\subsubsection{Combined effect of $J_\perp$ and $J_{\perp,2}$}
Based on the above, the two interlayer couplings compete with each other, since $J_\perp$ favours FM alignment of Cu2 dimers across the layers, whereas $J_{\perp,2}$ favours AFM alignment. 
When both couplings are taken into account, the numerical minimization of the classical energy  on large finite-size clusters, based on the {\it ab initio} values of Table~\ref{tab:abinitio} for $J_\perp$ and $J_{\perp,2}$, delivers a unique ground state. This is a coplanar state with eight spin sublattices in total, and with successive layers arranged antiparallel to each other. The configuration of each layer is uniform with four sublattices. 
Hence the combined effect of $J_\perp$ and $J_{\perp,2}$ is to lift the infinite degeneracy completely, but the selected state is {\it not} the one found experimentally.

We therefore conclude that the isotropic model of {\mat} is not enough to account for the experimentally reported state and that further couplings seem to play a qualitative role. In what follows we then turn to the investigation of the anisotropic interactions in {\mat}, starting from Dzyaloshinskii–Moriya anisotropy~\cite{Dzyaloshinsky1958, Moriya1960}.

\begin{figure}[!t]
\includegraphics[width=0.8\linewidth]{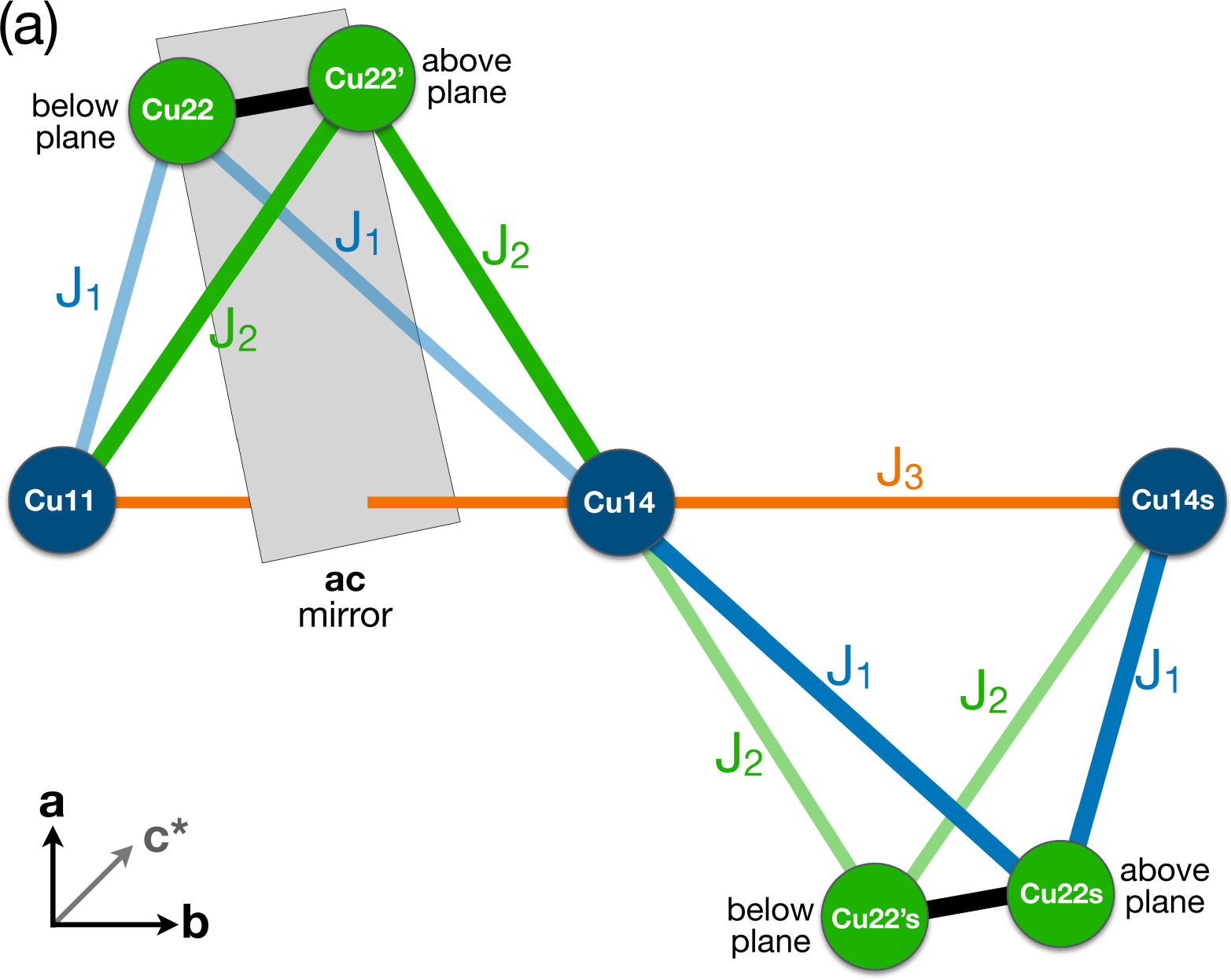}\\
\vspace*{0.2cm}
\includegraphics[width=0.8\linewidth]{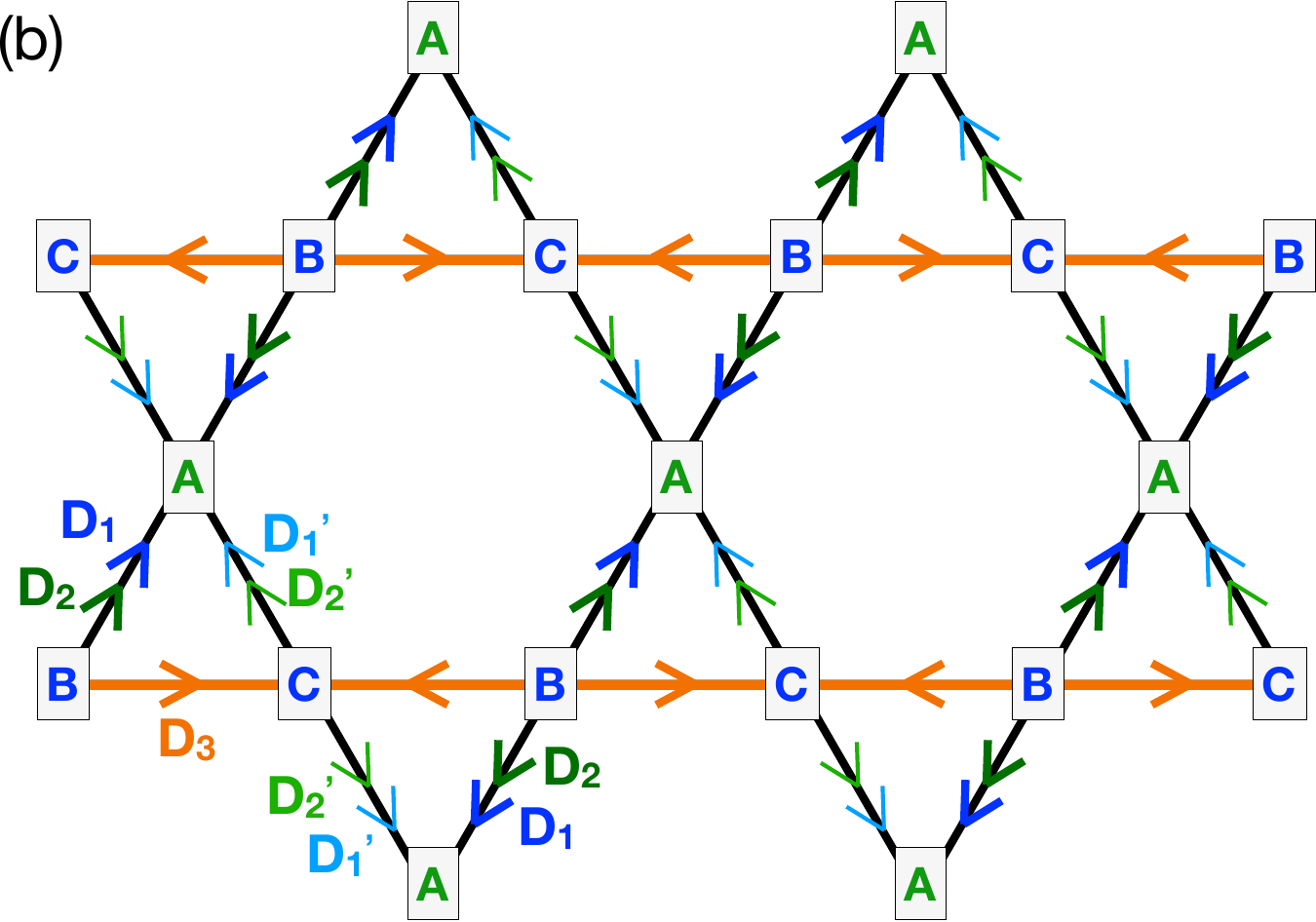}
\caption{(a) Local geometry used for the symmetry analysis of the DM vectors, see discussion in the main text. 
(b) Structure of DM vectors on the layers of {\mat}, where $J_1$ and $J_2$ bonds are projected onto each other. Arrows define the sequence of sites $(i,j)$ in the interaction term ${\bf D}_{ij}\cdot({\bf S}_i\times{\bf S}_j)$. The letters {\bf A}, {\bf B} and {\bf C} denote the uniform, 3-sublattice state found numerically (see text).}\label{fig:DMconstraints}
\end{figure}

\section{Dzyaloshinskii-Moriya Interactions}\label{sec:IncludingDM}

We now turn to the effect of Dzyaloshinskii-Moriya (DM) interactions, which enter the spin Hamiltonian in the general form
\be
\mc{H}_{\text{DM}} = \sum\nolimits_{i<j}
{\bf D}_{ij}\cdot \left( {\bf S}_i\times{\bf S}_j\right)\,,
\ee
where ${\bf D}_{ij}$ is the microscopic DM vector on the bond $(ij)$.

The presence of DM interactions in {\mat} has been discussed by Takahashi {\it et al}, who have also estimated an effective (coarse-grained) DM parameter $|{\bf D}|\simeq 7$\,K, from the value of the magnon gap measured by antiferromagnetic resonance~\cite{Takahashi2012}. Below, we show that this value underestimates significantly the actual size of the DM vectors because, i) not all of the DM components contribute to the magnon gap, and ii) the latter is affected by the symmetric part of the exchange anisotropy.

\subsection{Constraints from symmetry}\label{sec:DM-sym-constraints}
Let us first obtain the structure of the DM interactions using symmetry arguments. First of all, the inversion centers on the middle of the $J_d$ and $J_{\perp,2}$ bonds necessitate that the DM vectors on these bonds vanish identically. 
This leaves the DM vectors on the remaining bonds, $J_1$, $J_2$, $J_3$ and $J_\perp$. 

Let us consider the local geometry of two neighbouring building blocks shown in Fig.~\ref{fig:DMconstraints}\,(a), which shows three consecutive Cu1 sites, labeled as Cu11, Cu14, Cu14s, and four Cu2 sites (residing on the two adjacent Cu2 dimers), labeled as Cu22, Cu22', Cu22s and Cu22's.  
Now, the $ac^\ast$ mirror plane that crosses through the midpoint of $J_3$ bonds and contains the Cu2 dimers, maps $(S^a, S^b, S^{c^\ast})\to (-S^a, S^b, -S^{c^\ast})$ in spin space. Hence, the DM vectors connected by this operation are related to each other by a $\pi$-rotation. This implies, in particular, that the DM vectors on the $J_3$ bonds lie on the $ac^\ast$ plane.

Next, let us consider the constraints imposed by the screw axis along {\bf b}, which consists of a $C_{2b}$ rotation in spin-orbit space followed by a translation by ${\bf b}/2$. This operation maps Cu11$\to$Cu14, Cu14$\to$Cu14s, Cu22$\to$Cu22s, and Cu22’$\to$Cu22's in real space, and $(S^a, S^b, S^{c^\ast})\to(-S^a, S^b, -S^{c^\ast})$ in spin space. Hence, the DM vectors on any two bonds that are connected by this operation must also be related by a $\pi$-rotation around ${\bf b}$. 

Altogether, the symmetries impose the following inter-relations between the various DM vectors in Fig.~\ref{fig:DMconstraints}\,(a):
\be\renewcommand\arraystretch{1.2}
\begin{array}{ll}
\!{\bf D}_{11,14}\equiv {\bf D}_3,~ &\! {\bf D}_{14,14s}=-{\bf D}_3\\
\!{\bf D}_{11,22}={\bf D}_{14s,22s}\equiv{\bf D}_1,~ &\!{\bf D}_{14,22}={\bf D}_{14,22s}={\bf D}_1', \\
\!{\bf D}_{11,22'}={\bf D}_{14s,22's}\equiv {\bf D}_2,~ & \!{\bf D}_{14,22'}={\bf D}_{14,22's}= {\bf D}_2'\\
\end{array}
\ee
with $D_3^b=0$, and 
\be
{\bf D}_1' \equiv (-D_1^a,D_1^b,-D_1^{c^\ast}),~~ 
{\bf D}_2' \equiv(-D_2^a,D_2^b,-D_2^{c^\ast})\,. 
\ee
The above relations lead to a sign structure of the DM vectors shown in Fig.~\ref{fig:DMconstraints}\,(b). 

Similarly, the DM vectors, ${\bf D}_\perp$ and ${\bf D}_\perp'$, connecting two NN Cu1 sites with a Cu2 site on the next layer [see Fig.~\ref{fig:Structure})\,(b)]  are related to each other by the $ac^\ast$ mirror plane, and therefore 
\be
{\bf D}_\perp'=(-D_\perp^a,D_\perp^b,-D_\perp^{c^\ast})\,.
\ee

\subsubsection{Insights from cluster decomposition}
Similarly to what we did in Eq.~(\ref{eq:Hisoblocks1}), we can re-write the total DM energy as a sum over contributions from the building blocks of Fig.~\ref{fig:DMconstraints}\,(a), namely 
\be\label{eq:Hdmblocks1}
\mc{H}_{\text{DM}}=\sum\nolimits_{\parbox{0.25in}{\epsfig{file=UCell,width=0.2in,clip=}}} \!\!\!\mc{H}_{\text{DM},\parbox{0.25in}{\epsfig{file=UCell,width=0.2in,clip=}}}
\ee
with
\be\label{eq:Hdmblocks2}
\renewcommand\arraystretch{1.2}
\begin{array}{l}
\mc{H}_{\text{DM},\parbox{0.25in}{\epsfig{file=UCell,width=0.2in,clip=}}}
\!\!\!\!\!={\bf D}_3\cdot {\bf S}_1\times{\bf S}_2 
+{\bf D}_1\cdot {\bf S}_1\times{\bf S}_3
+{\bf D}_2\cdot {\bf S}_1\times{\bf S}_4\\
\hspace{1.5cm}
+{\bf D}_1'\cdot {\bf S}_2\times{\bf S}_3
+{\bf D}_2'\cdot {\bf S}_2\times{\bf S}_4\,.
\end{array}
\ee
As we shall see in Sec.~\ref{sec:JplusDM-GSs}, the classical ground state of the lattice model (with Heisenberg and DM couplings included) is a uniform, 3-sublattice state, similar to the state of Eq.~(\ref{eq:UCellState}), with all Cu2 sites parallel to each other, due to the strong $J_d$ coupling.
We can use this result and replace ${\bf S}_4\mapsto{\bf S}_3$ in (\ref{eq:Hdmblocks2}) to obtain the simpler expression
\be\label{eq:Hdmblocks2b}
\mc{H}'_{\text{DM},\parbox{0.25in}{\epsfig{file=UCell,width=0.2in,clip=}}}
\!\!\!\!\!={\bf D}_3\cdot {\bf S}_1\times{\bf S}_2 +
{\bf D}_{1,3}^{\text{eff}}\cdot {\bf S}_1\times{\bf S}_3+{\bf D}_{2,3}^{\text{eff}}\cdot {\bf S}_2\times{\bf S}_3.
\ee
where
\be\label{eq:Deff13and23a}
{\bf D}_{1,3}^{\text{eff}}={\bf D}_1+{\bf D}_2,~~~
{\bf D}_{2,3}^{\text{eff}}={\bf D}_1'+{\bf D}_2'.
\ee
One can also incorporate the contribution from the interlayer DM couplings by simply replacing these expressions with
\be\label{eq:Deff13and23b}
{\bf D}_{1,3}^{\text{eff}}={\bf D}_1+{\bf D}_2+{\bf D}_\perp,~~~
{\bf D}_{2,3}^{\text{eff}}={\bf D}_1'+{\bf D}_2'+{\bf D}_\perp',
\ee 
The above tell us that, due to the strong FM coupling $J_d$ on the Cu2 dimers, out of the six DM vectors ${\bf D}_1$, ${\bf D}_2$, ${\bf D}_1'$, ${\bf D}_2'$, ${\bf D}_\perp$ and ${\bf D}_\perp'$, only the combinations ${\bf D}_1+{\bf D}_2+{\bf D}_\perp$ and ${\bf D}_1'+{\bf D}_2'+{\bf D}_\perp'$ matter for the low-energy description of {\mat}, along with ${\bf D}_3$. As we show in the next subsection, when combined with our DFT results for the DM vectors, this is a key aspect for the understanding of the observed anisotropy in {\mat}.

\subsection{Insights from DFT calculations}\label{sec:DFT-DMvectors}
The DM vectors calculated using the \texttt{VASP} code are provided in the last column of Table~\ref{tab:abinitio}. The results show large DM vectors for all of the $J_1$, $J_2$, and $J_3$ bonds. The fact that these vectors are comparable in size to the Heisenberg terms seems quite unusual and can be traced back to the close competition between the $d_{x^2-y^2}$ and $d_{3z^2-r^2}$ magnetic orbitals. Microscopically, DM interactions can be understood as the effect of electron hoppings between the nonmagnetic (empty/filled) and magnetic (half-filled) $d$-orbitals in the presence of spin-orbit coupling~\cite{Moriya1960,Mazurenko2008}. By fitting the energy bands between $-1$\,eV and $+0.5$\,eV with a tight-binding model containing two orbitals ($d_{x^2-y^2}$ and $d_{3z^2-r^2}$) on each of the Cu sites, we find that such hoppings are comparable in size to hoppings between the magnetic orbitals. For example, we find $t_2=-0.136$\,eV for the hopping between $d_{x^2-y^2}$ of Cu1 and $d_{3z^2-r^2}$ of Cu2 (magnetic-to-magnetic) vs. $t_2'=0.131$\,eV for the hopping between $d_{x^2-y^2}$ of Cu1 and $d_{x^2-y^2}$ of Cu2 (magnetic-to-nonmagnetic). This would still lead to $|{\bf D}|/J<1$, but the Heisenberg term can be further lowered by a ferromagnetic contribution in the close-to-$90^{\circ}$ coupling geometry~\cite{Mazurenko2007}, and DM interactions can be eventually favored over isotropic exchange. 

Finally, besides the DM vectors on the $J_1$, $J_2$ and $J_3$ bonds, there also exists an interlayer DM interaction that lives on the $J_{\perp}$ bonds, which is however much weaker (see Table~\ref{tab:abinitio}) than the remaining anisotropies. 

There are two main insights from the DFT results. First, the strongest DM component is that of ${\bf D}_3$ along the ${\bf c}^\ast$ axis, $D_3^{c^\ast}=81$\,K. 
Second, the effective DM vectors connecting Cu1 and Cu2 sites are, according to Eq.~(\ref{eq:Deff13and23a}),  
\be\label{eq:D1p2a}
\!\!2\text{D}:~{\bf D}_{1,3}^{\text{eff}}=(12,-14,-68)\,\text{K},~
{\bf D}_{2,3}^{\text{eff}}=(-12,-14,68)\,\text{K},
\ee
or, according to Eq.~(\ref{eq:Deff13and23b}), which includes the effect of the interlayer DM vectors, 
\be\label{eq:D1p2b}
\!3\text{D}:~{\bf D}_{1,3}^{\text{eff}}=(17,-15,-71)\,\text{K},~
{\bf D}_{2,3}^{\text{eff}}=(17,-15,71)\,\text{K}.
\ee
Either way, these vectors have a very large component along ${\bf c}^\ast$ ($\pm68$\,K), like ${\bf D}_3$. Hence, we anticipate that the main effect of the DM interactions is to align the spins almost entirely on the $ab$ plane, which is consistent with experiment. 
Another effect is to increase the angle $2\theta$ between NN Cu1 spins beyond the value given by Eqs.~(\ref{eq:theta1b}) and (\ref{eq:theta1c}), as the DM favours a 90$^\circ$ orientation. 

Turning to the much weaker in-plane DM components, their role is to fix the orientation of the spin structure within the $ab$ plane and additionally give rise to a weak canting of the spins out of that plane. The specifics of these effects will be examined in Sec.~\ref{sec:JplusDM-GSs}.

\subsection{Classical ground state}\label{sec:JplusDM-GSs}

\subsubsection{Results from cluster decompositions and \\
unconstrained minimizations}

Let us return to the cluster decomposition of Eq.~(\ref{eq:Hdmblocks1}), and examine the ground state of $\mc{H}_{\text{DM},\parbox{0.25in}{\epsfig{file=UCell,width=0.2in,clip=}}}$. A numerical minimization of this four-site cluster delivers a ground state which is similar to that of Eq.~(\ref{eq:UCellState}) but with one crucial difference: the spins 3 and 4 of the Cu2 dimer show a weak misalignment. Due to the latter, this state {\it cannot} be tiled on the lattice and therefore the decomposition of Eq.~(\ref{eq:Hdmblocks1}) does not help to identify the exact ground state of the full lattice model. 
We then turn to the second smallest cluster, with two tetrahedral blocks,  
\be\label{eq:UCell3}
\includegraphics[width=1.25in]{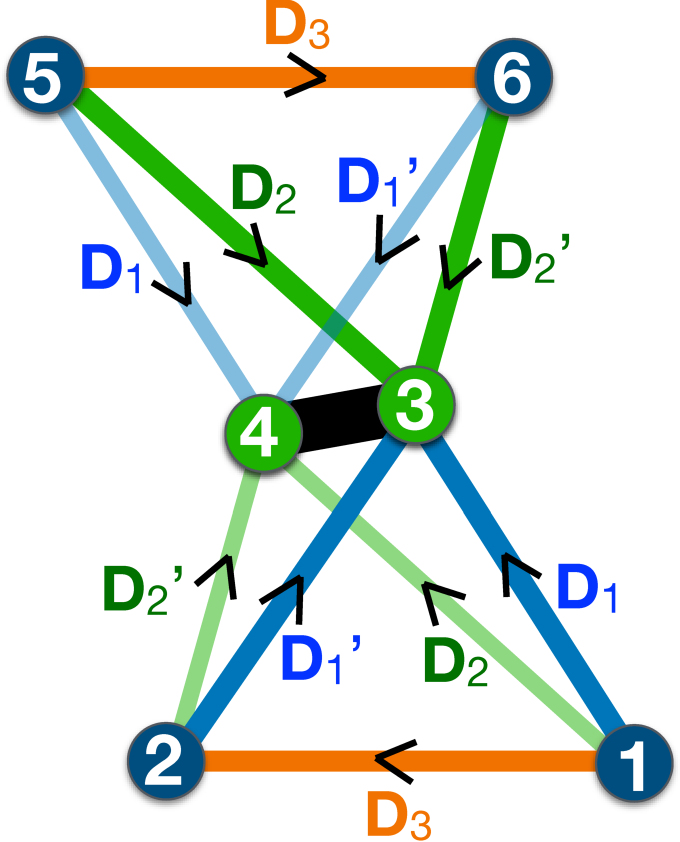}
\ee 
where we have also indicated the direction of the DM vectors on all bonds. The minimum energy configuration of this cluster shows no misalignment between the two spins of the Cu2 dimer (i.e., ${\bf S}_4={\bf S}_3$), and in addition ${\bf S}_5={\bf S}_1$ and ${\bf S}_6={\bf S}_2$. This state can be tiled on the lattice, leading to a uniform, three-sublattice state, as shown in Fig.~\ref{fig:DMconstraints}\,(b). This state coincides with the state we find independently via unconstrained numerical minimizations on large finite-size clusters with periodic boundary conditions.

The 3-sublattices ${\bf A}$, ${\bf B}$ and ${\bf C}$ of the above state can be depicted as
\be\label{eq:UniformState1}
\includegraphics[width=1.95in]{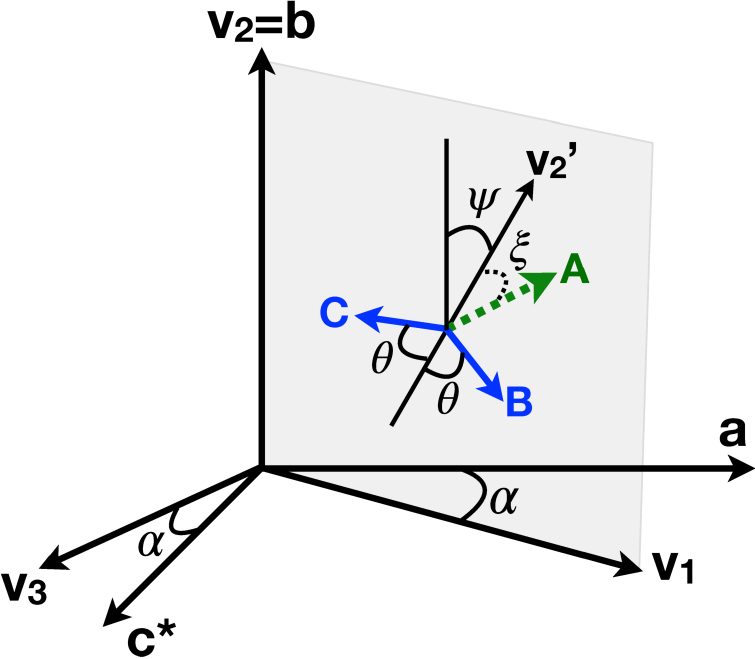}
\ee 
and their directions can be written as follows 
\be\label{eq:UniformState2}
\renewcommand\arraystretch{1.2}
\begin{array}{l}
{\bf S}_1/S=-\cos\theta~{\bf v}_2' + \sin\theta~{\bf v}_1' \equiv {\bf B},\\
{\bf S}_2/S=-\cos\theta~{\bf v}_2' - \sin\theta~{\bf v}_1' \equiv {\bf C},\\
{\bf S}_3/S={\bf S}_4/S=\cos\xi~{\bf v}_2' +\sin\xi~{\bf v}_3 \equiv {\bf A},
\end{array}
\ee
where 
\be\label{eq:UniformState3}
\renewcommand\arraystretch{1.2}
\begin{array}{c}
{\bf v}_1' = \cos\psi~{\bf v}_1 -\sin\psi~{\bf v}_2,\\
{\bf v}_2' = \sin\psi~{\bf v}_1 +\cos\psi~{\bf v}_2,\\
{\bf v}_1\!=\!\cos\alpha~\hat{\bf a}+\sin\alpha~\hat{\bf c}^\ast,\\
{\bf v}_2\!=\!\hat{\bf b},~~
{\bf v}_3\!=\!{\bf v}_1\times{\bf v}_2\,.
\end{array}
\ee
This configuration is qualitatively similar to the one of Eq.~(\ref{eq:theta1}). The angle $2\theta$ is the angle between the two Cu1 spins [as in Eq.~(\ref{eq:theta1})], the angle $\alpha$ specifies the plane of the Cu1 spins (${\bf S}_1$ and ${\bf S}_2$), the angle $\psi$ fixes the overall orientation within that plane, and the angle $\xi$ accounts for a small canting of the Cu2 spins (${\bf S}_3$ and ${\bf S}_4$) out of that plane. The state of Eq.~(\ref{eq:UniformState2}) reduces to that of (\ref{eq:theta1}) when $\xi\!=\!\psi\!=\!\alpha\!=\!0$, ${\bf v}_1\to{\bf e}_1$ and ${\bf v}_2\to{\bf e}_2$.

Let us discuss some further qualitative features of this state: 

i) The state is almost coplanar. The plane of the Cu1 spins lies very close to the $ab$-plane, and the Cu2 sites are tilted slightly away from that plane. Namely, the angles $\alpha$ and $\xi$ are both very small, see 6th and 8th column in Table~\ref{tab:models}. 
This can be traced back to the dominance of the $c^\ast$ components of ${\bf D}_3$, ${\bf D}_{1,3}^{\text{eff}}$ and ${\bf D}_{2,3}^{\text{eff}}$. Furthermore, our unconstrained numerical minimizations of large clusters shows that this result applies to both the single-layer and multiple-layer cases, with only small quantitative changes when the interlayer interactions are included (energy is lowered slightly, and angles between spins are renormalized). 

ii) The angle $2\theta$ between the two Cu1 spins is larger than the value given by Eqs.~(\ref{eq:theta1b}) and (\ref{eq:theta1c}), as the DM favours a 90$^\circ$ orientation, see fifth column in Table~\ref{tab:models}.
 
iii) The total moment per site is given by (assuming an isotropic ${\bf g}$ tensor, with $g=2$)
\be\label{eq:ClassMperCu2}
{\bf m}/ \mu_B = S [(\cos\xi-\cos\theta)~{\bf v}_2' + \sin\xi~{\bf v}_3]\,,
\ee
which reduces to Eq.~(\ref{eq:ClassMperCu}) for $\xi\!=\!0$ and ${\bf v}_2'\to{\bf e}_2$. The \texttt{VASP} values of Table~\ref{tab:abinitio} give a total moment per Cu site $m=0.22\mu_B$, for both the single layer and multilayer cases (see 4th column in  Table~\ref{tab:models}), which is very close to the experimental value~\cite{Zhao2019,Favre2020}.

iv) The direction of the total moment lies in the $ac^\ast$ plane, and predominantly  along the {\bf a}-axis (see 3rd column in  Table~\ref{tab:models}), and the angle $\psi\!=\!\pi/2$. The selection of this direction stems from the in-plane components of the DM vectors. This can be seen more directly by replacing the ansatz of Eq.~(\ref{eq:UniformState2}) into the expression for the energy of Eq.~(\ref{eq:Hdmblocks2b}), which reveals that the terms involving $\psi$ are all proportional to $\sin\psi$, and
\bea\label{eq:dEdpsi}
\frac{d}{d\psi} E'_{\text{DM},\parbox{0.25in}{\epsfig{file=UCell,width=0.2in,clip=}}} \!&\!=\!&\!
2 \cos\psi \sin\xi ~\Big\{\cos\theta~(D_{1,3}^{\text{eff}})^b \nonumber\\
&&\!\!\!- \sin\theta~\left[ 
(D_{1,3}^{\text{eff}})^a \cos\alpha + (D_{1,3}^{\text{eff}})^{c^\ast} \sin\alpha \right] 
\Big\},~~~
\eea 
and therefore the minimum energy corresponds to $\psi=\pm\pi/2$. 
This result disagrees with experimental data, which show that the total moment is along the ${\bf b}$ axis and not the ${\bf a}$ axis~\cite{Zhao2019,Favre2020}. 
Additionally, this disagrees with the refinement of neutron diffraction data, which delivers a state that belongs to the identity irrep $\Gamma_1$ $C2/m$ at zero momentum~\cite{Favre2020}, while the state favored by the in-plane DM components clearly breaks the $ac^\ast$ mirror plane and belongs to the $\Gamma_3$ irrep. 

So the model with Heisenberg and DM couplings included reproduces all experimental data except one, the direction of the uncompensated moment. According to Eq.~(\ref{eq:dEdpsi}), this deficiency cannot be remedied within the above model, unless we abandon the major insight from DFT on the separation of energy scales between the out-of-plane and the in-plane components of ${\bf D}_3$, ${\bf D}_{1,3}^{\text{eff}}$ and ${\bf D}_{2,3}^{\text{eff}}$. As this ingredient is crucial for getting the remaining aspects of the observed state right, the origin of the disagreement for the direction of the total moment must be sought in some other type of anisotropy, with the symmetric interactions ${\bf T}$ being the natural candidate.

The energy scale of this additional anisotropy should be at least as  strong as the in-plane components of ${\bf D}_3$, ${\bf D}_{1,3}^{\text{eff}}$ and ${\bf D}_{2,3}^{\text{eff}}$. To get an idea of this energy scale we have fixed the parameters $\alpha$, $\xi$ and $\theta$ to the values corresponding to the minimum energy configuration, and then tracked the variation of the total energy as we rotate $\psi$ away from $\pm\pi/2$. This calculation delivers a variational bound to the energy cost, which varies within a bandwidth of only $\sim 8.77$\,K as we vary $\psi$ between $0$ and $2\pi$. Hence, only a small amount of the additional anisotropy is necessary to rotate the moment along the direction found experimentally.

\section{Symmetric anisotropy ${\bf T}$}\label{sec:IncludingTbb}

The symmetric (and traceless) part of the anisotropy ${\bf T}$ enters the spin Hamiltonian in the form
\be
\mc{H}_{{T}} = \sum\nolimits_{i<j}
{\bf S}_i\cdot{\bf T}_{ij}\cdot{\bf S}_j\,,
\ee
where the $3\times3$ matrices ${\bf T}_{ij}$ have the properties ${T}_{ij}^{\alpha\beta}={T}_{ij}^{\beta\alpha}$ (where $\alpha$ and $\beta$ are Cartesian components) and $\text{Tr}({\bf T}_{ij})=0$.

\subsection{Constraints from symmetry}\label{sec:T-Sym-constraints}
The screw axis along ${\bf b}$ dictates that ${\bf T}_3$ is uniform along the chains (unlike the DM vector ${\bf D}_3$ which is staggered along the chains). Furthermore, the $ac^\ast$ mirror plane dictates that the only nonzero off-diagonal matrix elements of ${\bf T}_3$ are ${T}_3^{ac^\ast}={T}_3^{c^{\ast}a}$ and ${T}_3^{bc^\ast}={T}_3^{c^{\ast}b}$. The same is true for the matrix ${\bf T}_d$. 
By contrast, the matrix elements of ${\bf T}_1$ and ${\bf T}_2$ in the $(abc^\ast)$ frame are all nonzero, in principle. Furthermore, the screw axis along ${\bf b}$ dictates that the signs of the matrix elements ${T}_{1,2}^{ab}={T}_{1,2}^{ba}$ and ${T}_{1,2}^{c^{\ast}b}={T}_{1,2}^{bc^\ast}$ alternate from one building block of the structure to the next, as we travel along the Cu1 chains. For the bonds of Fig.~\ref{fig:DMconstraints}\,(a), we have, for example,
\be
{\bf T}_{11,22}\equiv{\bf T}_1=
\left(
\begin{array}{c c c}
     {T}_1^{aa} & {T}_1^{ab} & {T}_1^{ac^\ast} \\[10pt]
     {T}_1^{ab}& {T}_1^{bb} & {T}_1^{bc^\ast} \\[10pt]
     {T}_1^{ac^{\ast}} & {T}_1^{bc^\ast} & -{T}_1^{aa}-{T}_1^{bb}
\end{array}
\right)_{\{a,b,c^\ast\}}\,.
\ee
and 
\be
{\bf T}_{14,22}\equiv{\bf T}_1'=
\left(
\begin{array}{c c c}
     {T}_1^{aa} & -{T}_1^{ab} & {T}_1^{ac^\ast} \\[10pt]
     -{T}_1^{ab}& {T}_1^{bb} & -{T}_1^{bc^\ast} \\[10pt]
     {T}_1^{ac^{\ast}} & -{T}_1^{bc^\ast} & -{T}_1^{aa}-{T}_1^{bb}
\end{array}
\right)_{\{a,b,c^\ast\}}\,,
\ee
and similarly for ${\bf T}_{11,22'}\equiv{\bf T}_2$ and ${\bf T}_{14,22'}\equiv{\bf T}_2'$.

\subsection{Identifying the relevant element of the symmetric exchange tensor}\label{sec:Tmatrix-relevant}
Similarly to what we did in Eqs.~(\ref{eq:Hisoblocks1}) and (\ref{eq:Hdmblocks1}), we can re-write the total energy from ${\bf T}$ interactions as a sum over contributions from the building blocks of Fig.~\ref{fig:DMconstraints}\,(a), namely 
\be\label{eq:HGblocks1}
\mc{H}_{{T}}=\sum\nolimits_{\parbox{0.25in}{\epsfig{file=UCell,width=0.2in,clip=}}} \!\!\!\mc{H}_{{T},\parbox{0.25in}{\epsfig{file=UCell,width=0.2in,clip=}}}
\ee
with
\be\label{eq:HGblocks2}
\renewcommand\arraystretch{1.2}
\begin{array}{l}
\mc{H}_{{T},\parbox{0.25in}{\epsfig{file=UCell,width=0.2in,clip=}}}
\!\!\!\!\!= 
{\bf S}_3\cdot{\bf T}_d\cdot{\bf S}_4
+{\bf S}_1\cdot{\bf T}_3\cdot{\bf S}_2 
+ {\bf S}_1\cdot{\bf T}_1\cdot{\bf S}_3\\
\hspace{1.2cm}
+ {\bf S}_1\cdot{\bf T}_2\cdot{\bf S}_4
+ {\bf S}_2\cdot{\bf T}_1'\cdot{\bf S}_3
+ {\bf S}_2\cdot{\bf T}_2'\cdot{\bf S}_4\,.
\end{array}
\ee
Anticipating that the strong FM coupling $J_d$ will again enforce parallel (or almost parallel) alignment of the Cu2 spins, we can replace ${\bf S}_4\mapsto{\bf S}_3$ in (\ref{eq:HGblocks2}) to obtain the simpler expression
\be\label{eq:HGblocks3}
\mc{H}'_{{T},\parbox{0.25in}{\epsfig{file=UCell,width=0.2in,clip=}}}
\!\!\!\!\!=
{\bf S}_3\cdot{\bf T}_d\cdot{\bf S}_3+
{\bf S}_1\cdot{\bf T}_3\cdot{\bf S}_2 +
{\bf S}_1\cdot{\bf T}_{1,3}^{\text{eff}}\cdot{\bf S}_3
+{\bf S}_2\cdot{\bf T}_{2,3}^{\text{eff}}\cdot{\bf S}_3.
\ee
where ${\bf T}_{1,3}^{\text{eff}}={\bf T}_1+{\bf T}_2$ and ${\bf T}_{2,3}^{\text{eff}}={\bf T}_1'+{\bf T}_2'$. In principle, one can also add the contribution from the interlayer couplings (${\bf T}_\perp$, ${\bf T}_{\perp,2}$, and their mirror images ${\bf T}_\perp'$ and ${\bf T}_{\perp,2}'$), but we shall disregard them as they are likely much weaker compared to ${\bf T}_3$, ${\bf T}_d$, ${\bf T}_{1,3}^{\text{eff}}$ and ${\bf T}_{2,3}^{\text{eff}}$. 

In the following we shall set out to identify the components of the ${\bf T}$ matrices which favour a rotation of the total moment along the ${\bf b}$ axis, and as such can compete with the in-plane DM components.
We shall do this for the matrices ${\bf T}_d$ and ${\bf T}_3$ that correspond to the bonds with the strongest Heisenberg exchange paths, which are also the ones that lead to simple analytical insights.

\subsubsection{Effect of ${\bf T}_d$}
To understand the effect of ${\bf T}_d$, we examine its spectrum. The eigenvalues are given by
\be
\renewcommand\arraystretch{1.2}
\begin{array}{l}
\lambda_1^{(d)}\!=\!+{T}_d^{bb}\,,\\
\lambda_2^{(d)}\!=\!-\frac{{T}_d^{bb}}{2}-\!\sqrt{({T}_d^{bb}/2)^2+({T}_d^{aa})^2+({T}_d^{ac^\ast})^2+{T}_d^{aa}{T}_d^{bb}}\,,\\
\lambda_3^{(d)}\!=\!-\frac{{T}_d^{bb}}{2}+\!\sqrt{({T}_d^{bb}/2)^2+({T}_d^{aa})^2+({T}_d^{ac^\ast})^2+{T}_d^{aa}{T}_d^{bb}}\,,
\end{array}
\ee
and the corresponding eigenvectors by
\be
\renewcommand\arraystretch{1.2}
\begin{array}{l}
{\bf u}_1^{(d)}=\hat{{\bf b}}\,,\\
{\bf u}_2^{(d)}=({T}_d^{aa}+{T}_d^{bb}-\lambda^{(d)}_1)~\hat{\bf a}+{T}_d^{ac^\ast}~\hat{\bf c}^\ast\,,\\
{\bf u}_3^{(d)}=({T}_d^{aa}+{T}_d^{bb}-\lambda^{(d)}_2)~\hat{\bf a}+{T}_d^{ac^\ast}~\hat{\bf c}^\ast\,.
\end{array}
\ee
It follows that, if ${T}_d^{bb}\!<\!0$ then the ${\bf T}_d$ interaction favours alignment along the $\bf{b}$ axis, whereas if ${T}_d^{bb}\!>\!0$ it favours alignment in the $ac^\ast$ plane. So, the crucial ingredient is the sign of ${T}_d^{bb}$.

\subsubsection{Effect of ${\bf T}_3$}
By re-writing
\be
{\bf S}_1\cdot{\bf T}_3\cdot{\bf S}_2 = 1/2 ~\mc{S}^T\cdot \left(
\renewcommand\arraystretch{1.2}
\begin{array}{ll}
{\bf 0} & {\bf T}_3\\
{\bf T}_3 & {\bf 0}
\end{array}
\right)\cdot\mc{S},
\ee
where $\mc{S}=(S_1^a, S_1^b, S_1^{c^\ast}, S_2^a, S_2^b, S_2^{c^\ast})^T$, we see that one can understand the effect of ${\bf T}_3$ by examining the spectrum of the $6\times6$ matrix $\left(
\renewcommand\arraystretch{1.2}
\begin{array}{ll}
{\bf 0} & {\bf T}_3\\
{\bf T}_3 & {\bf 0}
\end{array}
\right)$. Now, ${\bf T}_3$ has the same form as ${\bf T}_d$, so its eigenvalues and eigenstates are, respectively, 
\be
\renewcommand\arraystretch{1.2}
\begin{array}{l}
\lambda_1^{(3)}\!=\!+{T}_3^{bb},\\
\lambda_2^{(3)}\!=\!-\frac{{T}_3^{bb}}{2}-\!\sqrt{({T}_3^{bb}/2)^2+({T}_3^{aa})^2+({T}_3^{ac^\ast})^2+{T}_3^{aa}{T}_3^{bb}},\\
\lambda_3^{(3)}\!=\!-\frac{{T}_3^{bb}}{2}+\!\sqrt{({T}_3^{bb}/2)^2+({T}_3^{aa})^2+({T}_3^{ac^\ast})^2+{T}_3^{aa}{T}_3^{bb}},
\end{array}
\ee
and
\be
\renewcommand\arraystretch{1.2}
\begin{array}{l}
{\bf u}_1^{(3)}=\hat{{\bf b}}\,,\\
{\bf u}_2^{(3)}=({T}_3^{aa}+{T}_3^{bb}-\lambda^{(3)}_1)~\hat{\bf a}+{T}_3^{ac^\ast}~\hat{\bf c}^\ast\,,\\
{\bf u}_3^{(3)}=({T}_3^{aa}+{T}_3^{bb}-\lambda^{(3)}_2)~\hat{\bf a}+{T}_3^{ac^\ast}~\hat{\bf c}^\ast\,.
\end{array}
\ee
The desired eigenvalues and eigenvectors of 
$\left(
\renewcommand\arraystretch{1.2}
\begin{array}{ll}
{\bf 0} & {\bf T}_3\\
{\bf T}_3 & {\bf 0}
\end{array}
\right)$ 
can then be expressed in terms of $\lambda_j^{(3)}$ and ${\bf u}^{(3)}_j$ (with $j=1-3$) as 
\be
\pm\lambda_j^{(3)}~~\text{and}~~
({\bf u}^{(3)}_j,\pm {\bf u}^{(3)}_j)^T\,,
\ee 
respectively. From these results, it follows that the crucial ingredient is the sign of the matrix element ${T}_3^{bb}$: If ${T}_3^{bb}\!>\!0$, the minimum eigenvalue is $\lambda_2^{(3)}$ and the form of the corresponding eigenvector $({\bf u}_2^{(3)},{\bf u}_2^{(3)})^T$ tells us that the total moment of the two Cu1 spins, ${\bf S}_1+{\bf S}_2$, must lie on the $ab$ plane.  
By contrast, if ${T}_3^{bb}\!<\!0$, the minimum eigenvalue is $\lambda_1^{(3)}$ and the form of the corresponding eigenvector $({\bf u}_1^{(3)},{\bf u}_1^{(3)})^T$ tells us that ${\bf S}_1+{\bf S}_2$ will be aligned with the ${\bf b}$ axis,  and will therefore compete with the in-plane components of the DM vectors. 
So, similarly to the case of ${\bf T}_d$, the most relevant aspect of ${\bf T}_3$ is the sign of ${T}_d^{bb}$.

\subsection{${\bf T}_3$ matrix from DFT \& effect on the ground state}\label{sec:DFT-Tbb}
Having identified the relevant matrix elements of the symmetric anisotropy, we have carried out DFT calculations based on the \texttt{VASP} code, in order to check if such matrix elements are in principle present in {\mat}, ii) have the right sign, and iii) are strong enough to compete with the in-plane DM components. The results for the $J_3$ bond,  
\be\label{eq:gamma9old}
{{\bf T}}_3=
\left(
\begin{array}{c c c}
     8.84 & 0 & 5.77 \\[10pt]
     0 & -5 & 0 \\[10pt]
     5.77 & 0 & -3.84
\end{array}
\right)_{\{a,b,c^\ast\}}\,,
\ee
(in unit of K), show that ${T}_3^{bb}$ is negative and of comparable size with the energy scale required to turn the moment along ${\bf b}$, as found experimentally~\cite{Zhao2019, Favre2020}. 

To check this explicitly, we have performed unconstrained numerical minimizations of a single layer of {\mat} using large clusters with periodic boundary conditions, and including the Heisenberg, DM and ${\bf T}_3$ couplings delivered by \texttt{VASP}. The results confirm that a ${T}_3^{bb}$ coupling of the order of $-5$\,K is enough to rotate the total moment along the ${\bf b}$ axis, as found experimentally~\cite{Zhao2019, Favre2020}. We have also found that the angle $\xi$ which describes the canting of the Cu2 spins out of the plane of the Cu1 spins, reduces to zero and the configuration becomes fully coplanar. Additionally, the angle $\alpha$ that describes the tilt of the spin plane away from the $ab$ plane decreases to 3-5$^\circ$. Such a small misalignment of the spin plane from the $ab$-plane can be easily missed experimentally. Finally, the total moment per Cu site is $0.21 \mu_B$ and $0.9 \mu_B$ for one and multiple layers, respectively, see Table~\ref{tab:models}. It is likely that the remaining ${\bf T}$ couplings will renormalize these numbers slightly, but the agreement to experiment~\cite{Zhao2019,Favre2020} is already very satisfactory.

\begin{figure}[t!]
\includegraphics[width=0.99\linewidth]{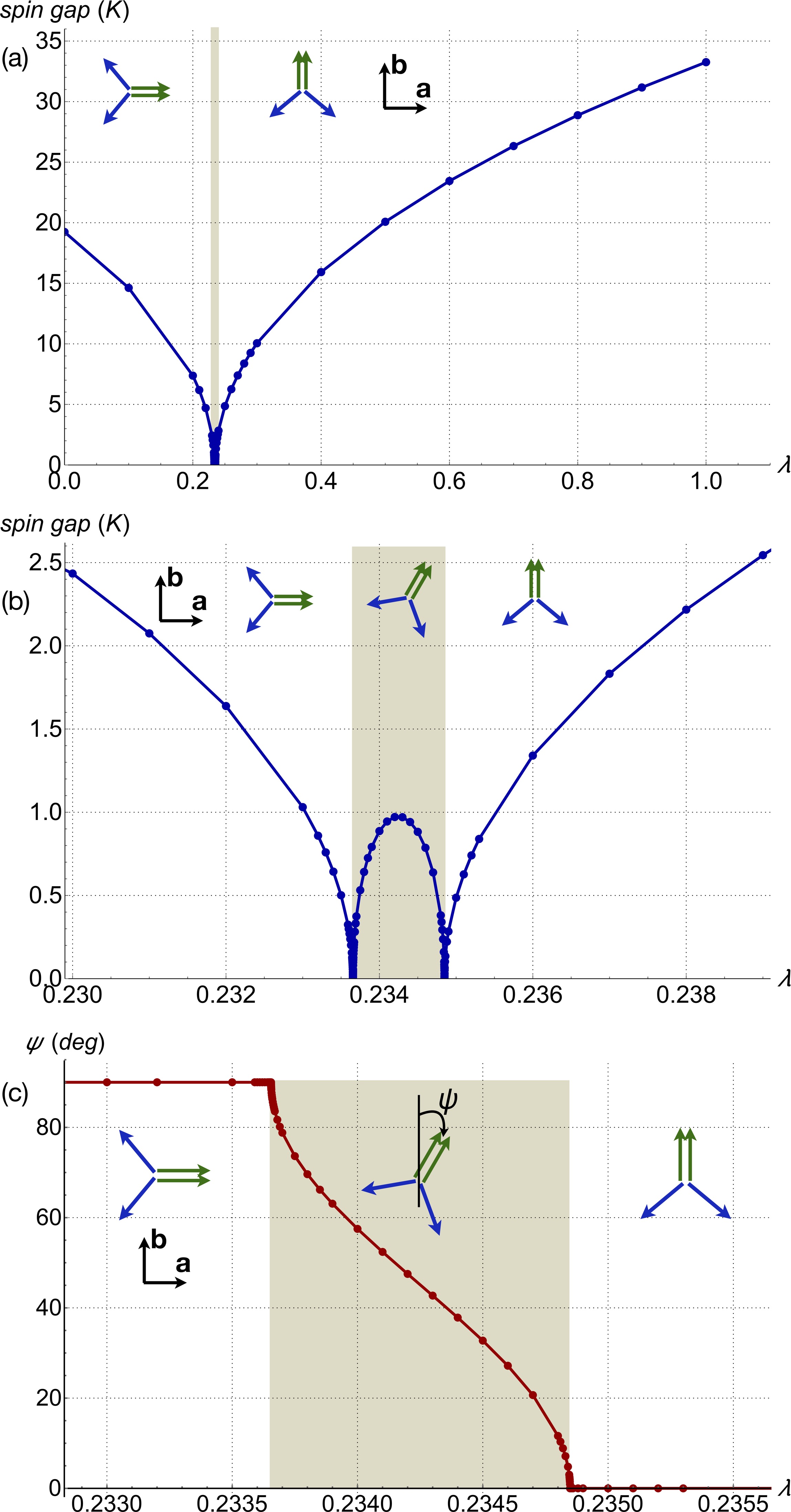} 
\caption{(a) Evolution of the spin gap with the rescaling parameter $\lambda$ of Eq.~(\ref{eq:lambda}), which characterizes the strength of the symmetric anisotropy ${\bf T}_3$. 
(b) Same as in (a), but zooming in the intermediate, re-orientation region. 
(c) Corresponding evolution of the angle $\psi$, see inset diagrams.}\label{fig:reorientation}
\end{figure}

\subsection{Nature of the re-orientation process \& the spin gap}\label{sec:reorientationtransition}

To uncover the nature of the re-orientation transition we examine the evolution of the classical ground state of the Hamiltonian 
\be\label{eq:lambda}
\mc{H}_\lambda=\mc{H}_{\text{iso}}+\mc{H}_{\text{DM}}+ \lambda~\mc{H}_{{T}_3}
\ee
as we vary the rescaling parameter $\lambda$ from zero to one. The results are summarized in Fig.~\ref{fig:reorientation}.
We find that the re-orientation proceeds via two continuous phase transitions, involving an intermediate phase (indicated by shading), where the total moment rotates from predominantly along the ${\bf a}$-axis to ${\bf b}$-axis. 
The boundary between the intermediate phase and the large-$\lambda$ phase involves the spontaneous breaking of the $ac^\ast$ mirror symmetry, whereas the boundary to the small-$\lambda$ phase involves the breaking of the symmetry $\mc{T}\times C_{2b}$, where $\mc{T}$ is the time reversal and $C_{2b}$ is the $\pi$-rotation around the $b$ axis in spin space alone; this combined operation flips the $b$ component of all spins.

{\it Spin gap --} The main curves of Fig.~\ref{fig:reorientation} show the evolution of the spin gap through the two-step re-orientation process, as obtained from linear spin-wave theory (see further discussion on magnon dispersions in Sec.~\ref{sec:LSWT} below). 
First of all, the value of the spin gap depends very sensitively on $\lambda$ across the two-step re-orientation transition. In particular, the spin gap drops to zero at the two boundaries of the intermediate phase. The closing of the gap at these boundaries is related to the spontaneous breaking of the symmetries discussed above.

Away from the intermediate phase, the spin gap reaches about 19~K and 33~K at $\lambda\!=\!0$ and $\lambda\!=\!1$, respectively. These values are set by  the energy scales of the in-plane DM components of ${\bf D}_3$, ${\bf D}_{1,3}^{\text{eff}}$ and ${\bf D}_{1,3}^{\text{eff}}$, and by that of ${\bf T}_3$ respectively. The out-of-plane DM components alone do not contribute to the spin gap because, in the absence of the in-plane DM components and ${\bf T}_3$, the system has a continuous U(1) symmetry and the states of Fig.~\ref{fig:reorientation} break this symmetry spontaneously. This is shown explicitly in Fig.~\ref{fig:DSF}\,(a) that we discuss in Sec.~\ref{sec:LSWT} below.

{\it Comparison to experiment --} Let us now compare the calculated values of the spin gap to the one extracted experimentally from antiferromagnetic resonance data, which is about $1.14$~K ~\cite{Takahashi2012}. 
This value is much smaller compared to the calculated gap at $\lambda\!=\!1$.  However, one should keep in mind that this value has been measured for a very weak mode that becomes visible below 5.3\,K only, i.e., at temperatures much lower than $T_N$. The nature of this mode must be clarified in future experiments. It is also worth noting that the above Hamiltonian includes the symmetric anisotropy on the $J_3$ bond only, and that the ${\bf T}$ couplings on the remaining bonds may compete with the effect of $T_3^{bb}$, thereby effectively reducing the spin gap and/or bringing {\mat} closer to the boundary with the intermediate phase, where the spin gap vanishes.

A related comment is in order here, which may connect the puzzle of the smallness of the observed spin gap with another puzzle that concerns the reported magnetization data for ${\bf H}\!\parallel\!{\bf a}$, see Fig.~5(b) of Ref.~\cite{Zhao2019}. According to these data, $m^a$ shows a small jump at zero field, from about $-0.0125$ to $+0.0125$ Bohr magnetons per Cu site. One plausible explanation can be the presence of an accidental misalignment of the field towards the ${\bf b}$ axis. Comparing the zero-field jump in $m^a$ for ${\bf H}\!\parallel\!{\bf a}$ to the corresponding jump in $m^b$ for ${\bf H}\!\parallel\!{\bf b}$, this misalignment would amount to only 3$^\circ$, which is a plausible explanation. 
However, an alternative scenario, which could also resolve the puzzle of the smallness of the spin gap,  is that the system may actually be slightly inside the intermediate phase. 
A definite answer to this requires further dedicated experimental studies.

\section{Other aspects and predictions for further experiments}\label{sec:OtherAspects}

\subsection{Magnetization process}\label{sec:MvsB}
To compare with reported magnetization curves we have also studied the effect of an applied external field ${\bf H}$ along the three directions ${\bf a}$, ${\bf b}$ and ${\bf c}^\ast$. The Zeeman coupling to the field takes the usual form, 
\be
\mc{H}_{\text{Z}}=-g\mu_B {\bf H} \cdot {\bf S}_{\text{tot}},~~~
{\bf S}_{\text{tot}}=\sum\nolimits_i {\bf S}_i\,,
\ee
where we have assumed an isotropic ${\bf g}$ tensor with $g=2$.

To find the evolution of the classical ground state of {\mat} for different field directions and field strengths, we have first performed unconstrained minimizations on large finite-size clusters with periodic boundary conditions. As it turns out, the state remains uniform for all field strengths. This allows us to decompose $\mc{H}_{\text{Z}}$ in terms of contributions from the clusters shown in (\ref{eq:UCell3}), namely 
\be\label{eq:Hz1}
\mc{H}_{\text{Z}}=
\sum_{\parbox{0.25in}{\epsfig{file=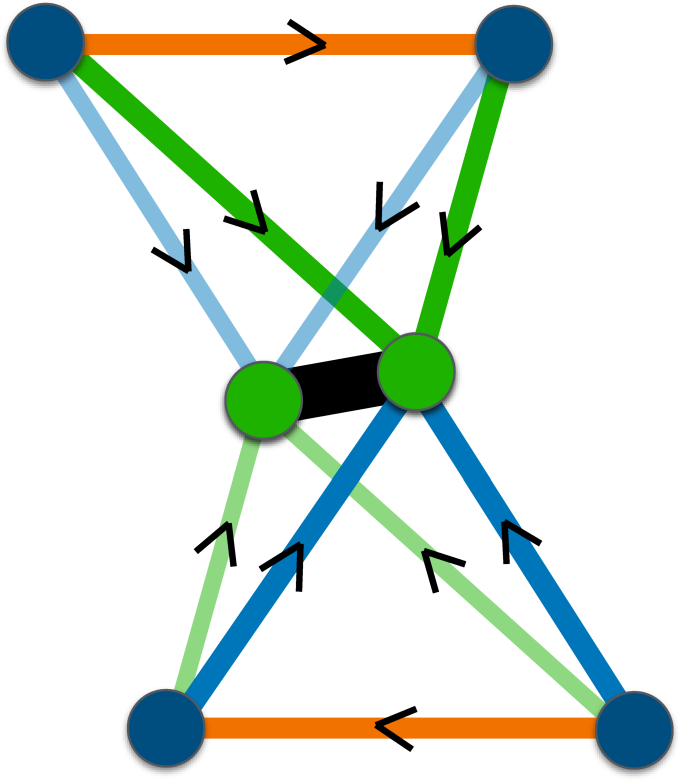,width=0.25in,clip=}}}
\mc{H}_{\text{Z},\parbox{0.25in}{\epsfig{file=UCell3b,width=0.25in,clip=}}}\,,
\ee
with
\be\label{eq:Hz2}
\mc{H}_{\text{Z},\parbox{0.25in}{\epsfig{file=UCell3b,width=0.25in,clip=}}}=-g\mu_B {\bf H}\cdot [\frac{1}{2}({\bf S}_1+{\bf S}_2+{\bf S}_5+{\bf S}_6)+{\bf S}_3+{\bf S}_4]\,,
\ee
where we use the site-labelling of (\ref{eq:UCell3}), and the prefactor of $1/2$ in the first term inside the square bracket takes care of double-counting. A numerical minimization of this 6-site cluster delivers, as before, ${\bf S}_4\!=\!{\bf S}_3$, ${\bf S}_5\!=\!{\bf S}_1$ and ${\bf S}_6\!=\!{\bf S}_2$, which simplifies Eq.~(\ref{eq:Hz1}) to   
\be\label{eq:Hz3}
\mc{H}_{\text{Z},\parbox{0.25in}{\epsfig{file=UCell3b,width=0.25in,clip=}}}=-g\mu_B {\bf H}\cdot ({\bf S}_1+{\bf S}_2+2{\bf S}_3)\,.
\ee

Before we present the numerical results, let us discuss some symmetry aspects. In the absence of a field, the system is invariant under the $ac^\ast$ mirror symmetry $\mc{M}_{ac^\ast}$, and also under the time reversal operation $\mc{T}$. 
Now, when the field is along the ${\bf a}$ or along the ${\bf c}^\ast$ axis, the system is invariant under the combined operation $\mc{M}_{ac^\ast} \times \mc{T}$, which maps
\be\label{eq:OpsHac}
\renewcommand\arraystretch{1.2}
\begin{array}{l}
(S_1^a, S_1^b, S_1^{c^\ast}) \rightarrow (S_2^a, -S_2^b, S_2^{c^\ast}),\\
(S_3^a, S_3^b, S_3^{c^\ast}) \rightarrow
(S_3^a, -S_3^b, S_3^{c^\ast})\,.
\end{array}
\ee
The classical ground states are found to break this symmetry spontaneously at a finite field $H^\ast$, below which the system shows the following nonzero order parameters: $S_1^a-S_2^a$, $S_1^b+S_2^b$, $S_1^{c^\ast}-S_2^{c^\ast}$ and $S_3^b$.

By contrast, in the presence of the field along the ${\bf b}$ axis, the $ac^\ast$ mirror plane survives, which means that the states related by the transformation
\be\label{eq:OpsHb}
\renewcommand\arraystretch{1.2}
\begin{array}{l}
(S_1^a, S_1^b, S_1^{c^\ast}) \leftrightarrow (-S_2^a, S_2^b, -S_2^{c^\ast}),\\
(S_3^a, S_3^b, S_3^{c^\ast}) \rightarrow
(-S_3^a, S_3^b, -S_3^{c^\ast}),
\end{array}
\ee
have the same energy. In principle, the classical ground states could break this symmetry spontaneously, but this is not supported by our numerics (the associated order parameters, which, in this case, are $S_1^a+S_2^a$, $S_1^b-S_2^b$, $S_1^{c^\ast}+S_2^{c^\ast}$, $S_3^a$ and $S_3^{c^\ast}$, all vanish irrespecive of the field strength).

\begin{figure*}[!t]
\includegraphics[width=0.99\linewidth]{MvsB2}
\caption{Evolution of the classical ground state of a single layer of {\mat} as a function of an external magnetic field applied along the ${\bf a}$ axis (a-c), ${\bf c}^\ast$ axis (d-f), and {\bf b} axis (g-i). The results are obtained by numerical minimizations on a single layer of {\mat}, using the Hamiltonian $\mc{H}_{\text{iso}}+\mc{H}_{\text{DM}}+\mc{H}_{T_3}+\mc{H}_{\text{Z}}$, see text. The first row of panels shows the three components of the total magnetic moment per Cu site, ${\bf m}$, in units of $\mu_B$. 
Panels (b) and (e) in the second row shows the combinations of spin components that play the role of order parameters, which set in at the fields $H^\ast_{a}$ and $H_c^{\ast}$. There is no such phase transition for fields along ${\bf b}$, see panel (h).
The last row shows the combination of spin components which are driven by the field, in a wider field range. The fields $H_a^{\ast\ast}$ and $H_b^{\ast\ast}$ correspond to abrupt re-orientation of the Cu1 spins, before we reach the saturation fields $H_{\text{sat}}$. There is no such transition for fields along ${\bf c}^\ast$.}
\label{fig:MvsB}
\end{figure*}

Our numerical minimization data shown in Fig.~\ref{fig:MvsB} give insights for the evolution of the classical ground state of a single layer of {\mat} as a function of an external magnetic field along ${\bf a}$ (panels a-c), {\bf c}$^\ast$ (d)-(f), and {\bf b} (g)-(i). The results for the multilayer case are qualitatively the same.  
The first row of panels shows the evolution of the three components of the total magnetic moment per Cu, ${\bf m
}$. 
The second row shows the combinations of Eq.~(\ref{eq:OpsHac}) and (\ref{eq:OpsHb}), which, as discussed above, play the role of order parameters. 
Finally, the last row shows the combination of spin components which are driven explicitly by the field and their evolution in a wider field range. 
Let us discuss the main features of Fig.~\ref{fig:MvsB}:

1) Unlike the case of a field along ${\bf b}$ (panel g), a field along ${\bf a}$ (panel a) and/or ${\bf c}^\ast$ (panel d), can induce appreciable transverse magnetization components, which can be detected experimentally by torque measurements. Such measurements can help to extract a more quantitative estimate of the relevant microscopic couplings.

2) Fields along ${\bf a}$ (panel b) and ${\bf c}^\ast$ (panel e) induce a phase transition at a characteristic field strength, $H_a^\ast$ and $H_c^\ast$, respectively. Below these fields, the symmetry $\mc{M}_{ac^\ast}\times\mc{T}$ is broken spontaneously, and the order parameters of Eq.~(\ref{eq:OpsHac}) become nonzero. 
By contrast, for fields along ${\bf b}$ (panel h), there is no such phase transition, and the mirror symmetry $\mc{M}_{ac^\ast}$ remains intact, irrespective of the field strength.

3) At the fields $H_a^\ast$ and $H_c^\ast$, the system has almost the same magnetic structure as in zero field, but with the total moment turned along the field direction. Here, the applied field works against the effect of ${\bf T}_3$ (which favours alignment along the ${\bf b}$ axis, as discussed above), and also partly against the in-plane components of the DM vectors (which favour alignment of the total moment in the $ac^\ast$ plane, and predominantly along the ${\bf a}$ axis). Along with the presence of multiple bonds, this can explain the large values obtained for $H_a^\ast$ and $H_c^\ast$.

4) The combinations of spin components that are driven explicitly by the field (data shown in the last row of panels in Fig.~\ref{fig:MvsB}) show that: i) the saturation field $H_{\text{sat}}$ is beyond 200~T for the microscopic parameters obtained from \texttt{VASP}, and ii) that the way to saturation in panels (c) and (i) involve an abrupt re-orientation of the Cu1 spins at some very high fields $H^{\ast\ast}_a$ and $H^{\ast\ast}_b$ below $H_{\text{sat}}$. Such a transition is absent for fields along ${\bf c}^\ast$.

5) According to the results shown in panels (a), (d) and (g) of Fig.~\ref{fig:MvsB}, the slopes of the longitudinal magnetizations (per Cu) at low fields are approximately 0.015~$\mu_B$/T for ${\bf H}\!\parallel\!{\bf a}$, 0.008~$\mu_B$/T for ${\bf H}\!\parallel\!{\bf c}^\ast$, and 0.0015~$\mu_B$/T for ${\bf H}\!\parallel\!{\bf b}$. The corresponding slopes estimated from the experimental data of Zhao {\it et al} at 2~K [Figure~5(b) of Ref.~\cite{Zhao2019}] are 0.015-0.016~$\mu_B$/T for ${\bf H}\!\parallel\!{\bf a}$,  0.012~$\mu_B$/T for ${\bf H}\!\parallel\!{\bf c}^\ast$, and 0.0028~$\mu_B$/T for ${\bf H}\!\parallel\!{\bf b}$. The agreement is satisfactory \footnote{We note that the data reported in Fig. 5(d) of Ref.~\cite{Zhao2019} must correspond to magnetic moment per Cu site and not per formula unit, to be consistent with the main finding of Ref.~\cite{Zhao2019} of a total moment of 1/4 $\mu_B$ per Cu ion, but also with Ref.~\cite{Favre2020}.}.

\section{Impact of quantum fluctuations}\label{sec:LSWT}
The picture presented so far has been entirely classical. The natural next step is to consider the effect of quantum fluctuations and confirm that their impact is minimal in {\mat}, as observed experimentally~\cite{Zhao2019,Favre2020}. 
As explained earlier, this aspect cannot be accounted for at the level of the isotropic model description of {\mat}, which remains highly frustrated despite the structural peculiarities of this material. 
The presence of symmetric and antisymmetric exchange anisotropy is crucial for resolving this puzzle as well, as these interactions open an appreciable spin gap (as shown already in Fig.~\ref{fig:reorientation}), which in turn alleviates the impact of quantum fluctuations. 
To show this explicitly we have performed a standard linear spin-wave expansion~\cite{Blaizot,White} around the ground state of a single layer of {\mat} and for various anisotropic terms included in the Hamiltonian. The multiple-layer problem does not feature any qualitative changes for the in-plane dispersion of the magnons and the onsite spin length reduction.

\subsubsection{Spin length reduction}\label{sec:SpinLength}
Let us take the site labelling convention of Fig.~\ref{fig:Structure}, and   denote by ${\bf w}_j$ (j=1-4) the direction of the four spins of the magnetic unit cell in the classical ground state; Incidentally, recall that these directions are different for ${\bf T}_3\!=\!0$ and for ${\bf T}_3\!\neq\!0$, as the corresponding ground states are different. 
The spin length of the $j$-th spin in the renormalized ground state is then given by $\langle S_j^w\rangle\!=\!S-\langle n_j\rangle$, where $S\!=\!1/2$ and $n_j$ is the number operator for the spin flips (magnons). 

For the model without ${\bf T}_3$ (and all DM components included) we find
\be
\text{DM only}:~~\Bigg\{
\renewcommand\arraystretch{1.2}
\begin{array}{l}
\langle{S}_1^w\rangle=\langle{S}_2^w\rangle =0.441,
\\
\langle{S}_3^w\rangle=\langle{S}_4^w\rangle=0.456, 
\end{array}
\ee
while for the model with both DM and ${\bf T}_3$ included we find
\be
\text{DM}+{\bf T}_3:~~\Bigg\{
\renewcommand\arraystretch{1.2}
\begin{array}{l}
\langle{S}_1^w\rangle=\langle{S}_2^w\rangle=0.437,
\\
\langle{S}_3^w\rangle=\langle{S}_4^w\rangle=0.451\,.
\end{array}
\ee
As expected by symmetry, the two Cu2 sites (j=3 and 4) have equal spin length and the same is true for the two Cu1 sites (j=1 and 2). Moreover, the effect of quantum fluctuations on the Cu2 sites is less severe due to the strong FM $J_d$ exchange. 
Most importantly, the reduction of the spin length is minimal (at most 12\% of the classical length of 1/2). This is consistent with experiment~\cite{Zhao2019,Favre2020}, and corroborates our earlier assertion that the impact of quantum fluctuations in {\mat} is heavily mitigated due to the appreciable anisotropy gap, giving further confidence to the above classical description.

\subsubsection{Magnon excitations \& dynamical structure factor}\label{sec:magnons}

We will now examine the spin-wave expansion more closely and discuss the magnon excitation spectrum. 
The uniform states discussed above have a four-site unit cell, and we therefore expect four magnon bands, emerging from the hybridization and delocalization of the four local spin flips on the unit cell.

\begin{figure}[t!]
\includegraphics[width=1\linewidth]{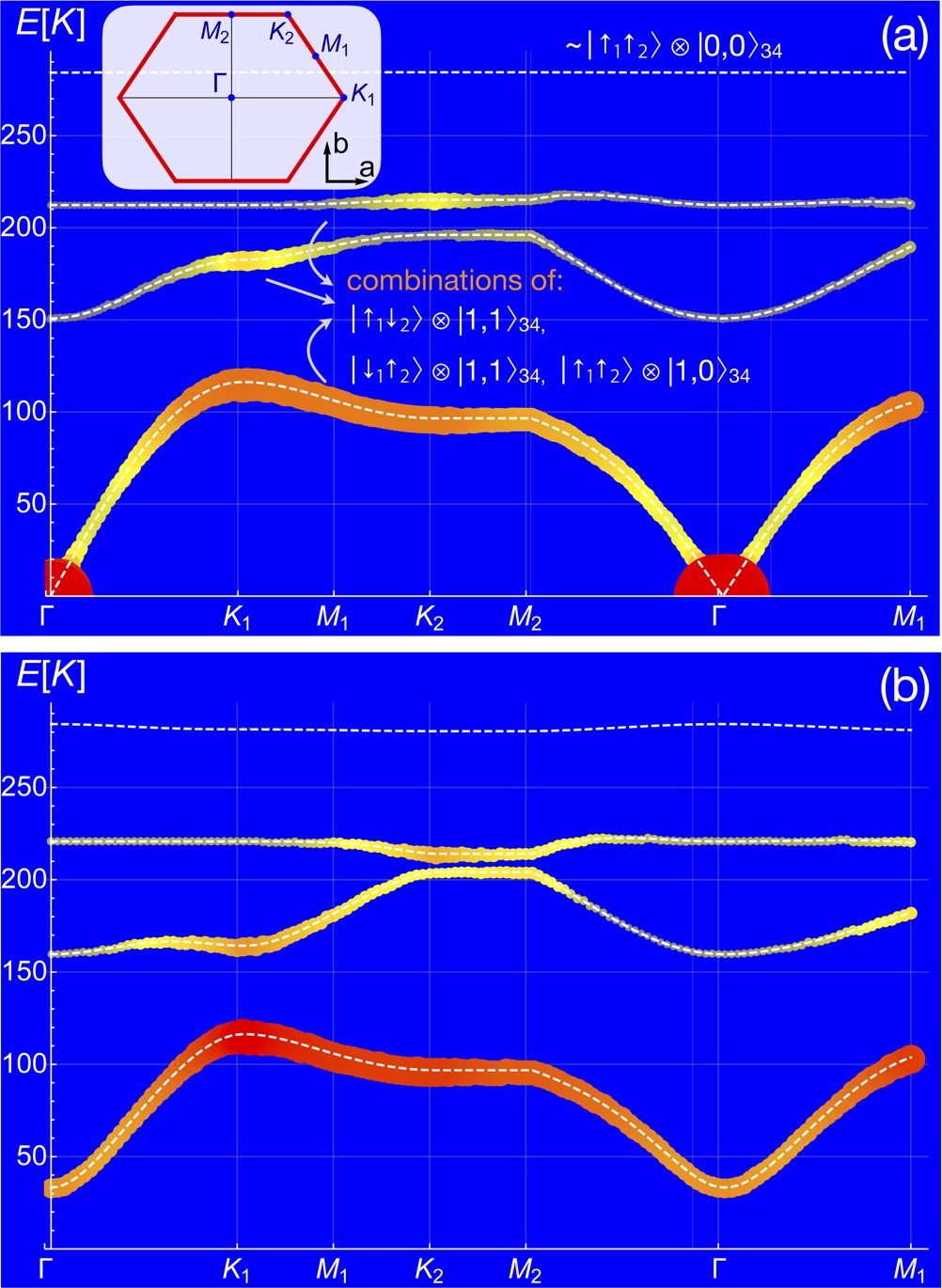}
\caption{Linear spin-wave spectra (white dashed lines) and dynamical structure factor intensity $\mc{S}({\bf Q},\omega)$ (symbol size, rescaled to the respective maximum intensity in each separate panel) for a single layer of {\mat}, along certain cuts of the Brillouin zone, see inset. In (a), the Hamiltonian includes Heisenberg couplings and the $c^\ast$ components of the DM couplings, whereas in (b) it includes Heisenberg, all components of the DM vectors, as well as the interaction matirx ${\bf T}_3$.}\label{fig:DSF}
\end{figure}

Figure~\ref{fig:DSF} shows the resulting magnon band structure of a single layer of {\mat}.   
The dispersions shown 
correspond to special cuts on the $ab$ plane, inside the first Brillouin zone (BZ), which is an almost symmetric hexagon (see inset). 
In panel (a), the spin Hamiltonian includes Heisenberg plus the $c^\ast$-components of the DM vectors, whereas in panel (b) it includes all DM components as well as the interaction matrix ${\bf T}_3$. In panel (a), the lowest mode is gapless at the $\Gamma$ point  because the Hamitlonian with only the $\bf{c}^\ast$-components of the DM vectors included has a U(1) symmetry around the ${\bf c}^\ast$ axis, and this symmetry is broken spontaneously by the classical ground state around which we expand. 
In panel (b), this mode is gapped out due to the in-plane DM components and ${\bf T}_3$, as discussed above. 
Apart from this qualitative difference, the remaining shape and overall positioning of the magnon bands is similar in the two panels, which reflects the fact that the intermediate- and high-energy parts of the excitation spectrum is not affected much by the anisotropies.

On top of the magnon bands of Fig.~\ref{fig:DSF} we also show by symbols (filled circles) the zero-temperature dynamical structure factor intensity (more specifically, its trace), 
\be
\mc{S}({\bf Q},\omega)=\int \frac{d\omega}{2\pi}~\Big{\langle
} {\bf M}({\bf Q},t)\cdot{\bf M}(-{\bf Q},0)\Big{\rangle}\,,
\ee
where ${\bf M}({\bf Q},t)$ is the time-evolved Fourier transform of the magnetization evaluated at wavevector ${\bf Q}$ and time $t$.
The size of the symbols (and colour) is scaled relatively to the maximum intensity in each separate panel.
The majority of the scattering weight is carried by the three lowest magnon branches. The fourth magnon band carries almost no weight at all, and features a much weaker dispersion with the bandwidth of the order of a few K.

\subsubsection{Further analysis of the magnon bands}
To gain some  understanding of the overall magnon band structure we proceed to a more in-depth analysis of the magnon modes. 
To that end, we take again the four-site labelling of the unit cell of (\ref{eq:UCell2}), and use, for each separate spin, the local quantization axis pertaining to the direction of that spin in the reference classical state around which we wish to expand. With this choice, the reference state is the tensor product of 
\be
|\!\uparrow_1\uparrow_2\rangle\otimes|S_{34}\!=\!1,M_{34}\!=\!1\rangle\,,
\ee 
over all unit cells, whre $S_{34}$ is the total spin quantum number of the Cu2 dimer and $M_{34}$ is its projection along the local quantization axis. At the mean-field level, the four magnon states and their corresponding excitation energies $\Delta{E}$ (measured from the reference state) are
\be\label{eq:magnoncontent1}
\renewcommand\arraystretch{1.25}
\begin{array}{l}
|\!\uparrow_1\uparrow_2\rangle\otimes|S_{34}\!=\!1,M_{34}\!=\!0\rangle\,,~~
\Delta{E}=h_3\,,\\
|\!\uparrow_1\downarrow_2\rangle\otimes|S_{34}\!=\!1,M_{34}\!=\!1\rangle\,,~~
\Delta{E}=h_1\,,\\
|\!\downarrow_1\uparrow_2\rangle\otimes|S_{34}\!=\!1,M_{34}\!=\!1\rangle\,,~~
\Delta{E}=h_1\,,\\
|\!\uparrow_1\uparrow_2\rangle\otimes|S_{34}\!=\!0,M_{34}\!=\!0\rangle\,,~~
\Delta{E}=|J_d|+h_3\,,
\end{array}
\ee
where $h_3\!=\!h_4$ is the magnitude of the local mean field exerted on a Cu2 site from its Cu1 neighbours, and, similarly, $h_1\!=\!h_2$ is the magnitude of the local mean field exerted on a Cu1 site from all its neighbours. 
Specifically, 
\be
\renewcommand\arraystretch{1.25}
\begin{array}{l}
{\bf h}_1=-2J_3\langle{\bf S}_2\rangle-2(J_1+J_2)\langle{\bf S}_3\rangle\,,\\
{\bf h}_3=-(J_1+J_2)\left(\langle{\bf S}_1\rangle+\langle{\bf S}_2\rangle\right)\,,
\end{array}
\ee
where $\langle{\bf S}_j\rangle$ (j=1-4) are the spin vectors in the classical reference state around which we expand. 
For the isotropic model, the magnitudes of these fields are given by
\be
h_1=J_3,~~~
h_3=\frac{(J_1+J_2)^2}{2J_3}\,,
\ee
whereas, in the presence of anisotropies, their values can be obtained numerically. 

The first member of Eq.~(\ref{eq:magnoncontent1}) is the symmetric combination of the two single spin-flips on the $J_d$ dimer, 
\be 
|S_{34}\!=\!1,M_{34}\!=\!0\rangle=\frac{1}{\sqrt{2}}
\Big(
|\!\uparrow_3\downarrow_4\rangle
+
|\!\downarrow_3\uparrow_4\rangle
\Big)\,,
\ee
whereas the last member of Eq.~(\ref{eq:magnoncontent1}) is the antisymmetric (singlet) combination,
\be 
|S=0,M=0\rangle_{3,4}=\frac{1}{\sqrt{2}}
\Big(
|\!\uparrow_3\downarrow_4\rangle
-
|\!\downarrow_3\uparrow_4\rangle
\Big)\,.
\ee

Returning to Fig.~\ref{fig:DSF}\,(a), it turns out the three lowest magnon branches arise, almost entirely, by the hybridization and delocalization of the first three members of Eq.~(\ref{eq:magnoncontent1}), whereas the fourth band, which resides at around $E=284$~K, is made almost entirely by the fourth member of Eq.~(\ref{eq:magnoncontent1}), i.e., the singlet.

The reason why this band is almost flat can be uncovered by examining what happens at the level of the isotropic model, i.e., by disregarding for the moment the effect of DM and ${\bf T}_3$ anisotropy. In that model, it can be shown that: i) a singlet excitation on the Cu2 dimer of one unit cell can hop (if $J_1\neq J_2$) to a Cu2 dimer of a neighbouring unit cell only via the intermediate Cu1 sites that connect these dimers, and ii) due to the geometry of the unit cell, the above hopping to the Cu1 sites is proportional to $J_1-J_2$. 
To see this, we simply start from the state $|\!\uparrow_1\uparrow_2\rangle\otimes|S_{34}\!=\!0,M_{34}\!=\!0\rangle$, and apply the relevant isotropic interactions $J_1$ and $J_2$, which can be grouped in the form
\be
\mc{H}_{\text{Cu1-Cu2}}=({\bf S}_1+{\bf S}_2)\cdot(J_1 {\bf S}_3+J_2{\bf S}_4)\,. 
\ee
One can then show that
\be
\renewcommand\arraystretch{1.2}
\begin{array}{l}
\mc{H}_{\text{Cu1-Cu2}} ~|S_{12}\!=\!1,M_{12}\!=\!1\rangle\otimes|S_{34}\!=\!0,M_{34}\!=\!0\rangle
\\
~~
=\frac{1}{2}(J_1-J_2) ~\Big\{
|S_{12}\!=\!1,M_{12}\!=\!1\rangle\otimes|S_{34}\!=\!1,M_{34}\!=\!0\rangle
\\
~~~
-|S_{12}\!=\!1,M_{12}\!=\!0\rangle\otimes|S_{34}\!=\!1,M_{34}\!=\!1\rangle
~\Big\}\,.
\end{array}
\ee
So the hopping of the excitation from the Cu2 dimer to the neighbouring Cu1 spins is indeed proportional to $J_1-J_2$. 
Now, according to the \texttt{VASP} values of Table~\ref{tab:abinitio}, this difference is small (8~K). So, the effective hopping from one Cu2 dimer to another is also small, which explains the very weak dispersion of the singlet branch; We have checked explicitly that, for $J_1=J_2$, the band becomes completely flat, as expected. 
Additionally, the small hopping amplitude gives a weak hybridization of this band with the remaining three bands, which explains our above statement that this band is made almost entirely of the singlet modes on the Cu2 dimers.  

Finally, we note that the $|S_{34}\!=\!1,M_{34}\!=\!-1\rangle$ member of the triplet on the Cu2 dimer corresponds, in the language of standard spin-wave expansion, to a two-particle excitation, which is why it does not appear in the spin wave spectrum of Fig.~\ref{fig:DSF}. A generalized, multiboson version of the spin-wave expansion, that incorporates the $J_d$ coupling explicitly on the Cu2 bonds, reveals this excitation branch in the spectrum around $\Delta{E}=2h_3$, which is approximately $232$~K at the level of the model $\mc{H}_{\lambda=1}$. The intensity $\mc{S}({\bf Q},\omega)$ of this mode vanishes, as it corresponds to a total $\Delta{M}\!=\!2$ excitation, relevant to the ground state. 
However, this mode can be observed, e.g., via magnetic Raman scattering that would then provide a more direct estimate of $2h_3$, and, in turn, of the microscopic couplings that affect $h_3$.

\section{Summary \& Discussion}\label{sec:Conclusions}

We have presented a comprehensive study of dolerophanite {\mat}, a layered kagome-like magnet where every
third site of the kagome structure is occupied by a strongly ferromagnetic spin dimer. 
We have arrived at a minimal microscopic model that uncovers the origin of most experimental data reported so far~\cite{Zhao2019,Favre2020}, including the intralayer and interlayer magnetic structure, the reported total magnetic moment, and the observed anisotropy in the direction of the uncompensated moment. 

Our results show that the origin of the canted order in {\mat} comes from antiferromagnetic NN interactions.  This fact sets this material apart from francisite and related compounds that feature canted order arising from \textit{ferromagnetic} NN interactions frustrated by an antiferromagnetic coupling between second neighbors~\cite{Rousochatzakis2015,Constable2017}. In Cu$^{2+}$ compounds, the sign of short-range exchange interactions is usually determined by the Cu--O--Cu angles in the crystal structure. The relevant values for {\mat} are $90.5/104.7^{\circ}$ ($J_1$), $117.6^{\circ}$ ($J_2$), and $114.2^{\circ}$ ($J_3$), all being similar to francisite and closer to $90^{\circ}$ than, e.g., in herbertsmithite ($\sim 119^{\circ}$) with its antiferromagnetic NN couplings. The antiferromagnetic nature of $J_1$, $J_2$, and $J_3$ is probably reinforced by the peculiar orbital configuration with the different symmetries of the magnetic orbital on the Cu1 and Cu2 sites, see Fig.~\ref{fig:superexchange}(d). 

Given the antiferromagnetic couplings in the kagome plane, the canted order can be broadly understood as a deformation of the $120^{\circ}$ state expected in KHAFM in the presence of perturbations, such as anisotropy and weak lattice distortions. This order may~\cite{Okuma2017} or may not~\cite{Iida2020} feature an uncompensated total moment depending on the exact symmetry of the material and the nature of perturbations therein. Such canted order in {\mat} is necessarily ferrimagnetic because one out of three sites is occupied by effective spin-1 moments of the Cu2 dimer. An interesting finding of our work is that the $\theta$-angle close to $60^{\circ}$ in {\mat}, i.e., an almost undistorted $120^{\circ}$ configuration, is a combined effect of geometrical frustration and anisotropy. Heisenberg exchange alone should result in much lower values of $\theta$ as a result of the sizable deformation of the kagome lattice. However, large anisotropy terms restore a close-to-$120^{\circ}$ spin configuration that allows a larger energy gain from the out-of-plane DM components. 

The large DM anisotropy in {\mat} is quite unusual and striking. It may be traced back to the close competition between $d_{x^2-y^2}$ and $d_{3z^2-r^2}$ as possible magnetic orbitals and the eventual stabilization of different orbitals on different Cu sites. A similar situation is realized in the skyrmion material Cu$_2$OSeO$_3$ where the sizable ratio of $|{\bf D}/J|\simeq 0.6$ was found for one of the couplings~\cite{Janson2014}. The simultaneous presence of $d_{x^2-y^2}$ and $d_{3z^2-r^2}$ magnetic orbitals may thus be a promising route for realizing large-DM interaction regime experimentally.  At any rate, a complete understanding of the precise mechanism behind the large DM interactions is still lacking, and warrants further dedicated investigations, e.g., via superexchange expansions that incorporate the precise low-symmetry crystal field of the two Cu sites, the various hopping amplitudes and the effect of the spin-orbit coupling. 

The presence of strong anisotropy is consistent with the fact that the observed magnetic moment is almost classical~\cite{Zhao2019,Favre2020}. This agrees with our linear spin-wave calculations, which demonstrate that the reduction of the spin length is very small in the presence of anisotropy. The latter ingredient is crucial, because the isotropic model description of {\mat} delivers an infinite ground state degeneracy at the classical level, which in turn is expected to give rise to large quantum fluctuations. 

Another key result is that the observed direction of the uncompensated moment results from a peculiar competition between the antisymmetric (Dzyaloshinskii–Moriya) and the symmetric part of the exchange anisotropy, which gives rise to a two-step re-orientation process involving two continuous phase transitions. 

One of the remaining open issues is the size of the magnon gap. Previous studies reported a small value of $1.14$~K using antiferromagnetic resonance~\cite{Takahashi2012}, but the exact nature of the mode probed in this experiment remains to be understood. To facilitate this understanding, we have presented predictions for further experiments, which can allow for a more quantitative refining of the microscopic coupling parameters. These include: i) the behaviour of {\mat} under magnetic fields along different crystallographic directions and the corresponding prediction of appreciable transverse magnetization that can be measured by torque experiments, ii) the prediction of the magnon excitation spectrum (and the associated dynamical spin structure factor) which can be measured by inelastic neutron scattering experiments, and iii) the prediction for the presence of a flat two-particle mode involving the  $|S_{34}\!=\!1,M_{34}\!=\!-1\rangle$ member of the triplet on the Cu2 dimers, with excitation energy $\Delta{E}\!=\!2h_3$, which can be measured by Raman scattering. 

We hope these predictions and the minimal model uncovered in this study will provide the basic framework and necessary guidelines for further theoretical and experimental works in {\mat} and related materials.

\begin{center}
{\bf Acknowledgements}
\end{center}
\vspace*{-0.1cm}
AT thanks Oleg Janson for his insightful remarks about the orbital state of Cu$^{2+}$ in dolerophanite. IR thanks Yang Yang, Natalia Perkins and Karlo Penc for fruitful discussions. 
The DFT computations for this work were done (in part) using resources of the Leipzig University Computing Center. 
This work has been supported by the Engineering and Physical Sciences Research Council, Grant No. EP/V038281/1, and by the Deutsche Forschungsgemeinschaft (DFG, German Research Foundation) -- TRR 360 -- 492547816. 
Additionally, we acknowledge the support by the National Science Foundation under Grant No. NSF PHY-1748958 and the hospitality of KITP, UC Santa Barbara, where part of this work was finalized.


\appendix

\section{Structural aspects}\label{app:StructuralAspects}
{\mat} crystallizes in a monoclinic crystal system with space group $C2/m$~\cite{Effenberger1985}. The lengths of the primitive translations along the three crystallographic axes are  $a=9.370\,{\AA}$, $b = 6.319\,{\AA}$ and $c= 7.639\,{\AA}$, and the angles between these axes are $\alpha=\gamma=\pi/2$ and $\beta=122.34^\circ$.
Table~\ref{tab:StructuralAspects} shows the actual positions of the Cu atoms mentioned in Table~\ref{tab:abinitio}.

\begin{table}
\caption{Positions of the Cu atoms mentioned in Table~\ref{tab:abinitio}, in fractional units (crystallographic coordinates) and in $\AA$ (Cartesian coordinates). In the third row from the end, the notation Cu23=Cu21p+${\bf a}$ +${\bf c}$ means that the displacement vector connecting the atoms labels as Cu23 and Cu21p is ${\bf a}+{\bf c}$.}
\begin{ruledtabular}
\begin{tabular}{lll}
    Cu site 		&  fractional units			&  	${\bf r}$~in~$abc^\ast$-frame (in ${\AA}$) \\
\toprule
Cu21 & (0.0721, 0, 0.2182) & ($-0.2161$, 0, 1.4083) \\
Cu21p & ($-0.0721$, 0, $-0.2182$) & (0.2161, 0, $-1.4083$) \\
Cu22 & (0.5721, 0.5, 0.2182) & (4.4689, 3.1595, 1.4083)\\
Cu22p &(0.4279, 0.5, $-0.2182$)  & (4.9011, 3.1595, $-1.4083$)\\
Cu23=Cu21p+{\bf a} +{\bf c} &(0.9279, 0, 0.7818)&  $({5.4997, 0, 5.0458})$\\
Cu11 & (0.25, 0.25, 0) & (2.3425, 1.57975, 0)\\
Cu14 & (0.25, 0.75, 0) & (2.3425, 4.73925, 0)
\end{tabular}
\end{ruledtabular}
\label{tab:StructuralAspects}
\end{table}

\section{Curie-Weiss temperature for anisotropic interactions}\label{appendix:CWT}
Here we provide a general formula for the Curie-Weiss temperature $\Theta_{\text{CW}}$ in a magnetic system with general, anisotropic interactions. 
To that end, we consider a spin Hamiltonian of the form 
\be
\mc{H}=\sum\nolimits_{i<j} {\bf S}_i\cdot{\bf K}_{ij}\cdot{\bf S}_j\,,
\ee
where ${\bf K}_{ij}$ is a $3\times 3$ matrix that embodies all possible bi-linear couplings between spins at sites $i$ and $j$. This coupling can be decomposed into the Heisenberg exchange $J_{ij}$, the DM anisotropy described by the DM vector ${\bf D}_{ij}$ and the symmetric and traceless part ${\bf T}_{ij}$ of ${\bf K}_{ij}$, as follows 
\be\label{eq:Kijdecompo}
{\bf K}_{ij} = J_{ij} {\bf 1} + {\bf A}_{ij}+{\bf T}_{ij}
\ee
where ${\bf 1}$ is the $3\times3$ identity matrix, 
\bea
&&
J_{ij}=\frac{1}{3}\text{Tr}\left({\bf K}_{ij}\right)\,,\\
&&
{\bf A}_{ij}=\frac{1}{2}
\left({\bf K}_{ij}-{\bf K}_{ji}\right),~~
D_{ij}^\alpha=\epsilon^{\alpha\beta\gamma}{\bf A}_{ij}^{\beta\gamma}\,,\\
&&
{\bf T}_{ij}=
\frac{1}{2}\left(
{\bf K}_{ij}+{\bf K}_{ji}\right)-J_{ij}{\bf 1},~~\text{Tr}\left({\bf T}_{ij}\right)=0\,,
\eea
$\epsilon^{\alpha\beta\gamma}$ is the Levi-Civita symbol, and $\alpha$, $\beta$ and $\gamma$ are Cartesian components.

Next, we would like to perform a high-$T$ expansion of the magnetic susceptibility and extract a formula for $\Theta_{\text{CW}}$, which, in general, will depend on the direction of the applied magnetic field ${\bf H}$. 
To that end, we use a mean-field decoupling, which is generally valid at high enough temperatures where the correlation length is very small. 
Due to the magnetic anisotropy, the direction of the (thermal averaged) field-induced moments $\langle {\bf S}_i\rangle$ are not necessarily the same. 
The total field exerted at site $i$, including the external field, is given by
\be
{\bf h}_i = -\frac{1}{g\mu_B}\sum\nolimits_j {\bf K}_{ij}\cdot\langle{\bf S}_j\rangle + {\bf H}\,,
\ee
where the minus sign in the first term on the right hand side follows from our convention for the Zeeman energy of site $i$, which is $-g\mu_B {\bf S}_i\cdot {\bf H}$. 
For spin-1/2, the thermal average $\langle {\bf S}_i\rangle$ in the mean-field decoupled problem, at temperature $T$, equals
\be\label{eq:MF1}
\langle {\bf S}_i\rangle= \frac{1}{2}\tanh\left(\frac{g\mu_B h_i}{2k_BT}\right)\hat{{\bf h}}_i
\ee
where $h_i=|{\bf h}_i|$ is the magnitude of the local field and $\hat{{\bf h}}_i$ is its direction. In the high-$T$ limit we can replace
\be
\langle {\bf S}_i\rangle =
\frac{g\mu_B}{4k_BT}~{\bf h}_i\,.
\ee
Using Eq.~(\ref{eq:MF1}) then gives
\be\label{eq:MF1b}
\langle S^\mu_i\rangle + 
r \sum\nolimits_{j\nu} K^{\mu\nu}_{ij}~\langle S^\nu_j\rangle = r ~g\mu_B~H^\mu\,,
\ee
where we have defined $r\equiv 1/(4k_BT)$. We note in passing that, for general spin $S$, this expression must be replaced with $r=S(S+1)/(3k_BT)$.  
With varying $i$ and $\mu$, Eq.~(\ref{eq:MF1b}) is a system of coupled equations which can be written in the compact form
\be\label{eq:MF2}
\bs{\Lambda}\cdot\bs{\sigma} = {\bf f}\,,
\ee
where, using the $abc^\ast$ frame, 
\be
{\bs{\sigma}}=
\renewcommand\arraystretch{1.25}
\begin{pmatrix}
\langle{S^a_1}\rangle \\
\langle{S^b_1}\rangle \\
\langle{S^{c^\ast}_1}\rangle \\
\langle{S^a_2}\rangle \\
\vdots \\
\langle{S^{c^\ast}_N}\rangle \\
\end{pmatrix},~~
{\bf f}=r ~g \mu_B 
\renewcommand\arraystretch{1.25}
\begin{pmatrix}
H^a \\
H^b \\
H^{c^\ast} \\
H^a \\
\vdots \\
H^{c^\ast}
\end{pmatrix}
\ee
and $\Lambda_{i\mu,j\nu}=\delta_{ij}\delta_{\mu\nu}+rK_{ij}^{\mu\nu}$, or, more compactly, ${\bs{\Lambda}}={\bf{1}}+r{\bf{K}}$. The solution of Eq.~(\ref{eq:MF2}) can be expanded in powers of $r$ (or, equivalently, in powers of $1/T$):
\be
    {\bs{\sigma}}={\bs{\Lambda}}^{-1}\cdot{\bs{f}}=({\bf{1}}+r{\bf{K}})^{-1}\cdot{\bf{f}}=({\bf{1}}-r{\bf{K}})\cdot{\bf{f}}+\mc{O}(r^3)
\ee
That is,
\be \langle{S^\mu_i}\rangle=r~g\mu_B \sum\nolimits_{\mu}\left(\delta_{\mu\nu}-r\sum\nolimits_{j}K_{ij}^{\mu\nu}\right)H^\nu+\mc{O}(r^3)\,.
\ee
Suppose now that we wish to obtain the diagonal elements of the susceptibility tensor. We apply the field in a fixed direction, say $\mu$, and calculate the total spin (per site) along the same direction:
\be    
\langle{S}^\mu_{\text{tot}}\rangle/N
= g\mu_B \left( r
-r^2\sum\nolimits_{ij}K_{ij}^{\mu\mu}/N\right)H+\mc{O}(r^3)\,,
\ee
from which we can read off the Curie-Weiss temperature by comparing to the high-$T$ expansion of the Curie-Weiss law
\be
\chi_{\mu\mu}=\frac{C}{T-\Theta_{\text{CW}}^{\mu\mu}}=\frac{C}{T}+\frac{C \Theta_{\text{CW}}^{\mu\mu}}{T^2}+\mc{O}(1/T^3)\,.
\ee
We find, for general $S$,
\be\label{eq:ThetaCWaniso}
\renewcommand\arraystretch{1.3}
\begin{array}{rl}
k_B\Theta_{\text{CW}}^{\mu\mu}=
&
-\frac{S(S+1)}{3}\frac{1}{N}\sum\nolimits_{ij}K_{ij}^{\mu\mu}\\
=&
-\frac{S(S+1)}{3}\frac{2}{N}\sum\nolimits_{i<j}(J_{ij}+T_{ij}^{\mu\mu})\,,
\end{array}
\ee
where in the second step we used Eq.~(\ref{eq:Kijdecompo}). Note that the DM interactions do not contribute to the Curie-Weiss temperature because of the antisymmetry of the tensor ${\bf A}_{ij}$.

Applying Eq.~(\ref{eq:ThetaCWaniso}) to the full model that we consider in the main text (which includes Heisenberg exchange, DM vectors and the symmetric anisotropy ${\bf T}_3$) gives
\be\label{eq:ThetaCWaniso2}
\renewcommand\arraystretch{1.25}
\!\!\begin{array}{l}
\Theta_{\text{CW}}^{aa} \!=\! -\frac{1}{8} [J_d\!+\!2(J_3\!+\!J_{\perp,2})\!+\!4(J_1\!+\!J_2\!+\!J_\perp) \!+\!2T_3^{aa}]\,,\\
\Theta_{\text{CW}}^{bb} \!=\! -\frac{1}{8} [J_d\!+\!2(J_3\!+\!J_{\perp,2})\!+\!4(J_1\!+\!J_2\!+\!J_\perp) \!+\!2T_3^{bb}]\,,\\
\Theta_{\text{CW}}^{c^\ast{c^\ast}} \!=\! -\frac{1}{8} [J_d\!+\!2(J_3\!+\!J_{\perp,2})\!+\!4(J_1\!+\!J_2\!+\!J_\perp)\!+\! 2T_3^{c^\ast{c^\ast}}]\,,
\end{array}
\ee
which reduces to Eq.~(\ref{eq:ThetaCW}) in the absence of ${\bf T}_3$.
\vspace{1cm}




%

\end{document}